\documentclass[journal,romanappendices]{IEEEtran}
\IEEEoverridecommandlockouts

\usepackage{algorithm,amsmath,amssymb,amsthm,bbm,cite,color,colortbl,graphicx,hhline,microtype,url,algpseudocode}
\usepackage[USenglish]{babel}
\usepackage{hyperref}
\usepackage[utf8]{inputenc}
\usepackage[T1]{fontenc}
\usepackage{multirow,makecell,tabularx}
\usepackage[figurename=Fig., labelsep=period, font=small]{caption}
\usepackage[nolist]{acronym}
\usepackage{dashrule}
\usepackage{textcase}

\usepackage{tikz,pgfplots}
\usetikzlibrary{arrows}
\pgfplotsset{compat=1.17}

\usepackage{titlesec}
\let\subparagraph\relax
\titlespacing{\section}{0pt}{6pt plus 2pt minus 1pt}{4pt plus 1pt minus 1pt} 
\titlespacing{\subsection}{0pt}{4pt plus 2pt minus 1pt}{2pt plus 1pt minus 1pt} 
\usepackage{array}
\newcolumntype{L}[1]{>{\raggedright\arraybackslash}m{#1}}
\newcolumntype{C}[1]{>{\centering\arraybackslash}m{#1}}

\renewcommand{\b}{\mathbf{b}}

\renewcommand{\d}{\mathbf{d}}

\newcommand{\f}{\mathbf{f}}

\newcommand{\n}{\mathbf{n}}

\newcommand{\p}{\mathbf{p}}
\newcommand{\q}{\mathbf{q}}
\renewcommand{\r}{\mathbf{r}}

\renewcommand{\u}{\mathbf{u}}

\newcommand{\w}{\mathbf{w}}
\newcommand{\x}{\mathbf{x}}
\newcommand{\y}{\mathbf{y}}
\newcommand{\z}{\mathbf{z}}

\newcommand{\0}{\mathbf{0}}
\newcommand{\1}{\mathbf{1}}

\newcommand{\C}{\mathbf{C}}

\newcommand{\F}{\mathbf{F}}
\newcommand{\G}{\mathbf{G}}
\renewcommand{\H}{\mathbf{H}}
\newcommand{\I}{\mathbf{I}}

\renewcommand{\P}{\mathbf{P}}

\newcommand{\V}{\mathbf{V}}
\newcommand{\W}{\mathbf{W}}

\newcommand{\mub}{\boldsymbol{\mu}}

\newcommand{\Sigmab}{\mathbf{\Sigma}}

\newcommand{\setC}{\mathcal{C}}

\newcommand{\setN}{\mathcal{N}}

\newcommand{\setS}{\mathcal{S}}

\newcommand{\Compl}{\mbox{$\mathbb{C}$}}
\newcommand{\Real}{\mbox{$\mathbb{R}$}}

\newcommand{\argmax}{\operatornamewithlimits{argmax}}
\newcommand{\argmin}{\operatornamewithlimits{argmin}}

\newcommand{\diag}{\mathrm{diag}}

\newcommand{\diff}{\mathrm{d}}
\newcommand{\Exp}{\mathbb{E}}
\newcommand{\herm}{\mathrm{H}}
\renewcommand{\Im}{\mathrm{Im}}
\newcommand{\logdet}{\mathrm{logdet}}

\newcommand{\minimize}{\mathrm{minimize}}
\renewcommand{\Pr}{\mathbb{P}}

\renewcommand{\Re}{\mathrm{Re}}
\newcommand{\sgn}{\mathrm{sgn}}

\newcommand{\tran}{\mathrm{T}}

\newcommand{\erf}{\mathrm{erf}}

\definecolor{oulu_blue}{HTML}{23408F}
\definecolor{oulu_green}{HTML}{39B54A}

\definecolor{red}{rgb}{1,0,0}
\definecolor{red_magenta}{rgb}{1,0,0.5}
\definecolor{magenta}{rgb}{1,0,1}
\definecolor{blue_magenta}{rgb}{0.5,0,1}
\definecolor{blue}{rgb}{0,0,1}
\definecolor{blue_cyan}{rgb}{0,0.5,1}
\definecolor{cyan}{rgb}{0,1,1}
\definecolor{green_cyan}{rgb}{0,1,0.5}
\definecolor{green}{rgb}{0,1,0}
\definecolor{green_yellow}{rgb}{0.5,1,0}
\definecolor{yellow}{rgb}{1,1,0}
\definecolor{red_yellow}{rgb}{1,0.5,0}
\begin{acronym}
\acro{ADC}{analog-to-digital converter}
\acro{AMP}{approximate message passing}
\acro{AWGN}{additive white Gaussian noise}
\acro{BLMMSE}{Bussgang LMMSE}
\acro{BLMMSE-DR}{Bussgang LMMSE with dither removal}
\acro{CSI}{channel state information}
\acro{D-BLMMSE}{doubly 1-bit quantized BLMMSE}
\acro{D-BLMMSE-DR}{doubly 1-bit quantized BLMMSE with dither removal}
\acro{D-ML}{doubly 1-bit quantized ML}
\acro{DAC}{digital-to-analog converter}
\acro{LMMSE}{linear MMSE}
\acro{MIMO}{multiple-input multiple-output}
\acro{ML}{maximum-likelihood}
\acro{ML-DR}{ML with dither removal}
\acro{MMSE}{minimum mean squared error}
\acro{MSE}{mean squared error}
\acro{QAM}{quadrature amplitude modulation}
\acro{SER}{symbol error rate}
\acro{SNR}{signal-to-noise ratio}
\acro{two-stage hoML}{two-stage homotopy optimization ML}
\acro{ULA}{uniform linear array}
\end{acronym}

\newcommand{\rx}{\textnormal{\tiny{RX}}}
\newcommand{\tx}{\textnormal{\tiny{TX}}}
\newcommand{\yy}{\textnormal{\tiny{Y}}}

\usetikzlibrary{arrows, arrows.meta, backgrounds, calc, intersections, decorations.markings, decorations.pathreplacing, positioning, shapes,fit}

\title{Data Detection for Massive MIMO Systems with 1-Bit Quantized Dithered Linear Precoding}
\author{Amin Radbord, Antti Tölli, and Italo Atzeni
\thanks{The authors are with the Centre for Wireless Communications, University of Oulu, Finland (e-mail: \{amin.radbord, antti.tolli, italo.atzeni\}@oulu.fi).}
\thanks{This work was presented in part at IEEE SPAWC 2025 \cite{Rad25}.}
\thanks{This work was supported by the Research Council of Finland (336449 Profi6, 348396 HIGH-6G, 357504 EETCAMD, and 369116 6G~Flagship) and by the Riitta and Jorma J. Takanen Foundation.}}

\begin{document}

\maketitle

\begin{abstract}
The power consumption of the analog-to-digital converters (ADCs) and digital-to-analog converters (DACs) in fully digital massive multiple-input multiple-output (MIMO) systems motivates the adoption of low-resolution architectures. In particular, 1-bit DACs reduce the power consumption and hardware complexity at the transmitter, but introduce severe transmit-side quantization distortion. In this paper, we investigate data detection for a point-to-point massive MIMO system with 1-bit DACs at the transmitter, where the linearly precoded signal is dithered prior to quantization, and either full-resolution or 1-bit ADCs at the receiver. Assuming that the dither vector applied at the transmitter is known at the receiver, we first develop soft-estimation-based data detection methods with symbol-independent dither removal for both full-resolution and 1-bit ADCs. We then introduce a new symbol-dependent linearization of the transmitted signal at the output of the 1-bit DACs and use it to derive maximum-likelihood (ML)-based data detection methods that directly recover the data symbol vector from the received signal. For full-resolution ADCs, this leads to an ML-based method with and without dither removal. For 1-bit ADCs, we develop an approximate ML-based method that exploits the derived statistics of the received signal without dither removal. We also propose low-complexity variants of the ML-based methods to mitigate the exponential complexity growth with the number of streams. Numerical results in terms of symbol error rate highlight the critical role of the dither power and demonstrate that the proposed ML-based methods (along with their low-complexity variants) achieve significant gains over a baseline based on binary ML detection via a homotopy algorithm.
\end{abstract}

\begin{IEEEkeywords}
1-bit ADCs, 1-bit DACs, dithering, massive MIMO, maximum-likelihood data detection.
\end{IEEEkeywords}

\section{Introduction} \label{sec:Intro}

Deploying massive \ac{MIMO} systems is essential to exploit the wide bandwidths available at high frequencies for 6G and future wireless networks \cite{Raj20}. To realize massive \ac{MIMO} arrays, fully digital architectures are particularly attractive since they provide highly flexible beamforming and large-scale spatial multiplexing without the need for the complex beam-management schemes of their hybrid analog-digital counterparts \cite{Atz25}. However, the power consumption of the \acp{ADC} and \acp{DAC} grows linearly with the bandwidth and exponentially with the number of resolution bits \cite{Atz21b}. This challenge has motivated extensive research on low-resolution \ac{ADC}/\ac{DAC} architectures. Among them, simple 1-bit \acp{ADC}/\acp{DAC} are especially appealing due to their minimal power consumption and complexity~\cite{Li17,Mol17,Atz23}, which makes large-scale deployment in massive arrays feasible. In particular, 1-bit \acp{DAC} relax the requirements for costly radio-frequency components at the transmitter and improve the energy efficiency by allowing the power amplifiers to operate without back-off.

\subsection{State of the Art} \label{sec:Inro.A}

There exists a vast literature on precoding design with 1-bit \acp{DAC}. Symbol-level precoding, which directly designs the 1-bit output exploiting constructive interference or constructive regions \cite{Sax17,AnLi17} or based on the \ac{MMSE} criterion \cite{Lian23}, can achieve strong performance at the expense of high computational complexity and symbol-dependent optimization. In contrast, quantized linear precoding applies linear precoding followed by independent 1-bit quantization, thereby offering a favorable trade-off between performance and complexity \cite{Jac17b}. For instance, \cite{Usm17} focused on linear precoding design for doubly 1-bit quantized systems, i.e., systems employing 1-bit \acp{DAC} and \acp{ADC}; however, its formulation relies on additional analog gains after quantization and involves several analytical approximations. Low-resolution transceiver architectures have also been investigated from energy-efficiency and information-theoretic perspectives. For instance, \cite{Cho22} proposed optimized precoding to maximize the energy efficiency in a multi-user downlink massive \ac{MIMO} systems with low-resolution \acp{ADC}/\acp{DAC}, while \cite{Xio21} analyzed the achievable rate of massive \ac{MIMO} relay systems with variable-resolution \acp{ADC}/\acp{DAC} with correlated fading. However, these works do not address direct data detection under transmit-side 1-bit quantization. To mitigate the quantization distortion, Gaussian dithering before the 1-bit \acp{DAC} was proposed in \cite{Sax19,Sax20}. These works showed that a suitable dither power can reduce the strong correlation among quantization distortion components across the transmit antennas and partially restore amplitude information, particularly for higher-order \ac{QAM} constellations. More broadly, the benefits of dithering are widely recognized in low-resolution quantized systems \cite{Rap19}, including massive \ac{MIMO} systems with 1-bit \acp{ADC} \cite{Atz22}.

While 1-bit \acp{DAC} reduce power consumption and hardware complexity at the transmitter, they also introduce severe transmit-side quantization distortion, which motivates developing receive-side data detection methods that compensate for this distortion. In this paper, we focus on a point-to-point massive \ac{MIMO} system with 1-bit \acp{DAC} and either full-resolution or 1-bit \acp{ADC} at the receiver. This setting is particularly relevant because the receiver can mitigate the transmit-side distortion by processing the signal received across its many antennas. Despite the advances in precoding design, direct receive-side data detection that accounts for 1-bit \ac{DAC} quantization at the transmitter remains unexplored for point-to-point massive \ac{MIMO}. In general, data detection methods for 1-bit quantized systems have been developed primarily for multi-user uplink massive \ac{MIMO} systems with 1-bit \acp{ADC} at the receiver. In this context, \cite{Cho16} introduced a two-stage near \ac{ML} data detection method, where the first stage identifies a subset of candidates for the subsequent stage. The work in \cite{Ngu21} considered an approximate \ac{ML} data detection problem and solved it via deep-unfolded gradient descent, while \cite{San24} proposed a gradient-descent-based approach applied to a smooth surrogate of the likelihood function of the 1-bit quantized signals. Moreover, \cite{Saf24} reduced the \ac{ML} search space through a two-step framework exploiting the Hessian of the log-likelihood objective. The work in \cite{Rad26} proposed soft-estimation-based data detection techniques based on the statistical properties of the soft-estimated symbols. However, all these methods are tailored to multi-user uplink systems with 1-bit \acp{ADC} and therefore do not directly apply to the point-to-point massive \ac{MIMO} setting considered here. In our case, 1-bit quantization takes place at the transmitter before channel propagation and additive noise, which leads to a fundamentally different data detection problem.

Since no existing data detection method directly matches the considered system model, a natural baseline consists in first detecting the binary transmit signal at the output of the 1-bit \acp{DAC} and then recovering the original data symbols in a subsequent \ac{ML}-based data detection stage. This motivates considering binary data detection methods as a starting point. Among them, \cite{Sha21} introduced binary \ac{ML} detection based on a homotopy algorithm for multi-user uplink massive \ac{MIMO} systems with either full-resolution or 1-bit \acp{ADC} under quadrature phase-shift keying signaling, whose real-valued representation yields a binary transmit vector. The homotopy algorithm traces a solution path from a convex relaxation to the original nonconvex binary \ac{ML} problem, thereby reducing the risk of convergence to poor local minima. This method was shown to outperform approaches based on sphere decoding \cite{Dam03}, semidefinite relaxation \cite{Wai11}, and \ac{AMP} \cite{Ran19} in systems with full-resolution \acp{ADC}, as well as generalized \ac{AMP} \cite{Wan14,Wen16} and the two-stage near \ac{ML} data detection method from \cite{Cho16} with 1-bit \acp{ADC}. However, the formulation in \cite{Sha21} ultimately addresses binary \ac{ML} detection at the receiver and does not account for the effect of 1-bit \acp{DAC} at the transmitter on the recovery of phase and amplitude information. Therefore, applying it to our setting would require an additional \ac{ML}-based data detection stage after detecting the output of the \acp{DAC}, which introduces error propagation between the two stages.

\subsection{Contribution} \label{sec:Intro.B}

In this paper, we consider a point-to-point massive \ac{MIMO} system with 1-bit \acp{DAC} and either full-resolution or 1-bit \acp{ADC} at the receiver. Unlike \cite{Sax19,Sax20}, which optimized and evaluated dithering for multi-user downlink transmission with single-antenna users, we focus on a distinct multi-stream point-to-point massive \ac{MIMO} setting in which the transmitter employs linear precoding and dithering followed by 1-bit quantization. In this setting, dithering is not only a transmit-side design parameter but also directly shapes the statistics of the received signal and the resulting data detection problem. This motivates the development of data detection methods that explicitly account for dithering at the transmitter. In this context, we propose new soft-estimation-based and \ac{ML}-based data detection methods for systems with full-resolution and 1-bit \acp{ADC}. For soft-estimation-based data detection, we derive new combining matrices with dither removal based on the \ac{LMMSE} criterion. For \ac{ML}-based data detection, we develop new methods both with and without dither removal. The proposed methods capture different trade-offs between performance and computational complexity under distinct \ac{ADC} resolutions and operating regimes.

The contributions of this paper are summarized as follows:
\begin{itemize}
\item Assuming that the dither vector applied at the transmitter is known at the receiver, so that dither removal can be performed, we propose soft-estimation-based data detection methods with symbol-independent dither removal, namely \textit{\ac{BLMMSE-DR}} and \textit{\ac{D-BLMMSE-DR}} for full-resolution and 1-bit \acp{ADC}, respectively.

\item We introduce a new symbol-dependent linearization of the transmitted signal at the output of the 1-bit \acp{DAC}. Based on this, we develop an \ac{ML}-based data detection method for full-resolution \acp{ADC} that directly detects the data symbol vector from the received signal. We begin by presenting this method without dither removal, deriving the full likelihood function of the received signal along with its statistics. We then incorporate symbol-dependent dither removal, which yields the \textit{\ac{ML-DR}} method. For data detection with 1-bit \acp{ADC}, we develop an approximate \ac{ML}-based method, referred to as \textit{\ac{D-ML}}, which exploits the previously derived statistics of the received signal without dither removal. For both proposed data detection methods, we also develop low-complexity variants to mitigate the exponential complexity growth with the number of streams.

\item Numerical results demonstrate the effectiveness of the proposed methods in terms of \ac{SER}, also in comparison with a baseline referred to as \ac{two-stage hoML}, whose first stage employs the homotopy algorithm in \cite{Sha21}. First, the results highlight the critical role of the dither power: an optimal operating point appears only at moderate-to-high \ac{SNR}, while all the methods benefit from dither removal for moderate-to-large dither powers. Second, thanks to dither removal, the soft-estimation-based methods can even outperform the \ac{two-stage hoML} baseline despite their significantly lower complexity. Third, the proposed \ac{ML}-based methods provide the largest gains: \ac{ML-DR} and \ac{D-ML}, together with their low-complexity variants, consistently outperform both the soft-estimation-based methods and \ac{two-stage hoML} for full-resolution and 1-bit \acp{ADC}, respectively.
\end{itemize}
Part of this work was presented in our conference paper \cite{Rad25}, which derived the likelihood function of the received signal along with its statistics and proposed the \ac{ML}-based data detection method without dither removal for the case of full-resolution \acp{ADC}.

\smallskip

\textit{Outline.} The rest of the paper is organized as follows. Section~\ref{sec:Sys} introduces the system model. Section~\ref{sec:LQP&DD} presents linearization approaches for the 1-bit quantized transmitted and received signals. Sections~\ref{sec:DD1} and~\ref{sec:DD2} constitute the core technical contribution and propose new data detection methods with full-resolution and 1-bit \acp{ADC}, respectively. Section~\ref{sec:CCA&NR} presents the complexity analysis and numerical results. Lastly, Section~\ref{sec:Contr_Full} concludes the paper.

\smallskip

\textit{Notation.} Boldface lowercase and uppercase letters represent vectors and matrices, respectively, whereas calligraphic letters denote sets. $(\cdot)^\tran$, $(\cdot)^\herm$, and $(\cdot)^*$ represent the transpose, Hermitian transpose, and conjugate operators, respectively. $\I_N$ denotes the $N$-dimensional identity matrix, whereas $\0_N$ and $\1_N$ represent the $N$-dimensional all-zero and all-one vectors, respectively. $[\cdot]_{m,n}$ specifies the $(m,n)$th element of the matrix argument, whereas $x_{n}$ and $x_{\textrm{a},n}$ denote the $n$th elements of $\x$ and $\x_{\textrm{a}}$, respectively. $\| \cdot \|_{2}$ denotes the Euclidean norm. $\diag(\cdot)$ produces a diagonal matrix with the diagonal elements of the square matrix argument. $\Re[\cdot]$ and $\Im[\cdot]$ denote the real and imaginary parts, respectively, whereas $j = \sqrt{-1}$ is the imaginary unit. $\mathcal{CN}(\0_N,\Sigmab)$ and $\mathcal{N}(\0_N,\Sigmab)$ are, respectively, the $N$-variate complex and real normal distributions with zero mean and covariance matrix $\Sigmab$. $\sgn(\cdot)$ is the sign function. $\mathbb{E}[\cdot]$ represents the expectation operator, $\mathbb{P}[\cdot]$ denotes probability, and $\{\cdot\}$ is used to denote sets. $\erf(x) \triangleq \frac{2}{\sqrt{\pi}} \int_{0}^{x} e^{-t^2} \diff t$ represents the error function for $x \in \Real$; for $x \in \Compl$, we have $\erf_{\textrm{c}}(x) \triangleq \erf \big(\Re[x] \big) +j \, \erf \big(\Im[x] \big)$. $\Phi(x) \triangleq \frac{1}{2}\big(1 + \erf(\frac{x}{2})\big)$ represents the normal cumulative distribution function for $x \in \Real$. Lastly, $\mathrm{A}[\cdot]$ and $\mathrm{B}[\cdot]$ can be either $\Re[\cdot]$ or $\Im[\cdot]$, whereas $\Lambda_{\mathrm{A},\mathrm{B}}(\cdot): \Compl^{\ell_1 \times \ell_2} \to \mathbb{R}^{\ell_1 \times \ell_2}$ denotes the elementwise operator defined as $\Lambda_{\mathrm{A},\mathrm{B}}(\cdot) \triangleq \delta_{\mathrm{A},\mathrm{B}} \Re[\cdot] - (1-\delta_{\mathrm{A},\mathrm{B}})\Im[\cdot]$, with $\delta_{\mathrm{A},\mathrm{B}} = 1$ if $\mathrm{A} = \mathrm{B}$ and $\delta_{\mathrm{A},\mathrm{B}} = 0$ otherwise.

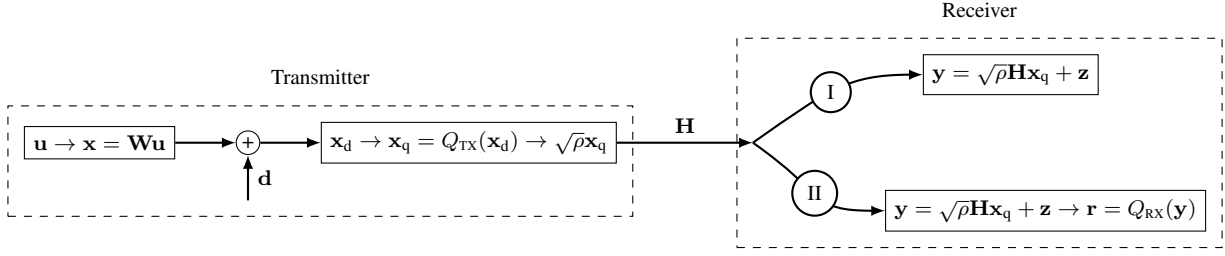
\begin{figure*}[t!]
\centering
\begin{tikzpicture}[>=latex, font=\footnotesize, node distance=2cm]

\node[draw, rectangle, align=center] (sx)
{$\u \rightarrow \x = \W\u$};

\node[circle, draw, inner sep=1pt, right=0.8cm of sx] (sum) {+};

\draw[->, black, thick] ([yshift=-0.6cm]sum.south) -- node[right] {\textcolor{black}{$\mathbf{d}$}} (sum.south);

\coordinate (ditherTip) at ([yshift=-0.6cm]sum.south);

\node[draw, rectangle, align=center, right=0.8cm of sum] (tx)
{$\x_{\textrm{d}} \rightarrow \x_{\textrm{q}} = Q_{\tx}(\x_{\textrm{d}}) \rightarrow \sqrt{\rho}\x_{\textrm{q}}$};

\node[draw, dashed, fit=(sx)(tx)(ditherTip), inner sep=6pt] (txbox) {};
\node[above=0.15cm of txbox] {Transmitter};

\draw[->, thick] (sx.east) -- (sum.west);
\draw[->, thick] (sum.east) -- (tx.west);

\node[right=1.8cm of tx] (y) {};

\draw[->, thick] (tx.east) -- node[midway, above] {$\H$} (y.west);

\node[draw, rectangle, right=2cm of y, yshift=9mm, align=center] (rx_top)
{$\y = \sqrt{\rho}\H\x_{\textrm{q}} + \z$};

\node[draw, rectangle, right=1.5cm of y, yshift=-9mm, align=center] (y_block)
{$\y = \sqrt{\rho}\H\x_{\textrm{q}} + \z \rightarrow \r = Q_{\rx}(\y)$};

\draw[->, thick] (y.west) .. controls ++(1.0,0.7) and ++(-1.0,0) ..
(rx_top.west) node[pos=0.45, circle, draw, fill=white] {I};

\draw[->, thick] (y.west) .. controls ++(1.0,-0.7) and ++(-1.0,0) ..
(y_block.west) node[pos=0.45, circle, draw, fill=white] {II};

\node[draw, dashed, fit=(y)(rx_top)(y_block), inner sep=6pt] (rxbox) {};
\node[above=0.15cm of rxbox] {Receiver};

\end{tikzpicture}
\caption{Diagram of the considered (doubly) 1-bit quantized massive \ac{MIMO} system.}\label{fig:diagram}
\end{figure*}

\section{System Model} \label{sec:Sys}

Consider a point-to-point massive \ac{MIMO} system where a transmitter equipped with $N$ antennas and 1-bit \acp{DAC} transmits $K$ data streams to a receiver with $M$ antennas, with $K \leq \min(N, M)$. The receiver employs either full-resolution or 1-bit \acp{ADC}, as detailed in the following. This setup may represent, for instance, a wireless backhaul scenario, as considered in Section~\ref{sec:NR}. To model the 1-bit \acp{DAC} and \acp{ADC}, we introduce the elementwise 1-bit quantization function $Q_{\textnormal{\tiny{Y}}} (\cdot) : \Compl^{a} \to \sqrt{\frac{\eta_{\textnormal{\tiny{Y}}}}{2}} \{ \pm 1 \pm j \}^{a}$, with
\begin{align}
Q_{\yy} (\b) \triangleq \sqrt{\frac{\eta_{\yy}}{2}} \Big( \sgn \big( \Re [\b] \big) + j \, \sgn \big( \Im [\b] \big) \Big) \label{eq:Q},
\end{align}
where the output is a vector of quadrature phase-shift keying symbols scaled by $\eta_{\yy} > 0$. In the following, we use the subscripts $\textnormal{Y} = \textnormal{TX}$ and $\textnormal{Y} = \textnormal{RX}$ to indicate ``transmitter'' and ``receiver'', respectively.

Let \( \H \in \mathbb{C}^{M \times N} \) represent the channel matrix between the transmitter and receiver. In this work, we assume that perfect \ac{CSI} is available at both the transmitter and receiver. This assumption is reasonable for point-to-point massive \ac{MIMO} systems with no mobility and quasi-static propagation environment, such as backhaul links between two fixed base stations. In such settings, the channel varies slowly over time and can be accurately estimated using sufficiently long pilots. While the main analysis focuses on this setting, the impact of imperfect \ac{CSI} at the receiver is also evaluated through numerical results in Section~\ref{sec:NR}; a detailed analysis under imperfect \ac{CSI} is left for future work.

The transmitter aims at delivering the data symbol vector $\u \triangleq [u_{1}, \ldots, u_{K}]^{\tran} \in \Compl^{K}$ to the receiver. We consider $\Exp[\u] = \0_{K}$ and $\u \in \setS^{K}$, where $\setS \triangleq \{ s_{1}, \ldots, s_{L} \}$ represents the transmit constellation with $L$ data symbols. Without loss of generality, we assume that $\setS$ corresponds to the 16-\ac{QAM} constellation, i.e., $\setS = \frac{1}{\sqrt{10}} \{ \pm 1 \pm j, \pm 1 \pm j \, 3, \pm 3 \pm j, \pm 3 \pm j \, 3 \}$, which is normalized to ensure $\frac{1}{L} \sum_{l=1}^{L} |s_{l}|^{2} = 1$; nonetheless, the proposed framework is general and can be applied to any transmit constellation. The transmitter employs a 1-bit quantized linear precoding strategy with Gaussian dithering, as in \cite{Sax19,Sax20}. First, the precoding matrix $\W \in \Compl^{N \times K}$ is computed based on $\H$ and independently of $\u$. Then, the resulting signal $\x \triangleq \W \u \in \Compl^{N}$ is supplemented with the Gaussian dither vector $\d \sim \mathcal{CN}(\0_{N}, \sigma^2 \I_N)$ with dither power $\sigma^2$, which helps mitigate the effects of coarse quantization arising in the subsequent 1-bit quantization step. Specifically, Gaussian dithering can reduce the strong correlation among the quantization distortion components across the transmit antennas, particularly when the number of data streams is small. Given the dithered precoded signal $\x_{\textrm{d}}\triangleq \x + \d \in \mathbb{C}^{N}$, the transmitted signal after the 1-bit \acp{DAC} is given by
\begin{align}\label{eq:xq}
    \x_{\textrm{q}} \triangleq Q_{\tx}(\x_{\textrm{d}}) \in \Compl^{N},
\end{align}
where the scaling factor of the 1-bit quantization function in \eqref{eq:Q} is fixed to $\eta_{\tx} = \frac{1}{N}$ to satisfy the power constraint $\|\x_{\textrm{q}}\|_2^2 = 1$. In Section~\ref{sec:LQP&DD}, we introduce two linearization approaches to linearize the transmitted signal $\x_{\textrm{q}}$ with respect to the dithered precoded signal $\x_{\textrm{d}}$.

Subsequently, the analog signal $\x_{\textrm{q}}$ is transmitted over the channel with transmit power $\rho$. From this point onward, we consider two alternative receiver settings:  
\begin{itemize}
\item[(I)] \textbf{Full-resolution \acp{ADC}.} The signal observed at the receiver is given by
\begin{align}\label{eq:y}
    \y & \triangleq \sqrt{\rho}\H \x_{\textrm{q}} + \z \in \mathbb{C}^{M},
\end{align}
where $\z \sim \mathcal{CN}(\0_{M}, \I_M)$ is a vector of \ac{AWGN}. Since the \ac{AWGN} is assumed to have unit variance, $\rho$ can be interpreted as the transmit \ac{SNR}.

\item[(II)] \textbf{1-bit \acp{ADC}.} The signal observed at the receiver is given by
\begin{align} \label{eq:r}
\r \triangleq Q_{\rx}(\y) \in \Compl^{M},
\end{align}
which consists of $\y$ in \eqref{eq:y} after passing through the 1-bit ADCs. Here, the scaling factor of the 1-bit quantization function in \eqref{eq:Q} is fixed to $\eta_{\rx} = \rho + 1$, ensuring that the variance of $\r$ matches that of $\y$.
\end{itemize}
The considered (doubly) 1-bit quantized massive \ac{MIMO} system is illustrated in Fig.~\ref{fig:diagram}. The observed signals in \eqref{eq:y} or \eqref{eq:r} are used for data detection, either through a soft-estimation step or via \ac{ML}-based methods (which process the full observed signal). In the former case, a soft estimate $\check{\u} \in \mathbb{C}^{K}$ of the data symbol vector is first obtained via linear combining as
\begin{align} \label{eq:hat_s}
    \check{\u} \triangleq
    \begin{cases}
        \V^\herm \y & \textrm{for (I)}, \\
        \V^\herm \r & \textrm{for (II)},
    \end{cases}
\end{align}
where $\V \in \mathbb{C}^{M \times K}$ is the combining matrix at the receiver. Each soft-estimated symbol in \eqref{eq:hat_s} is then mapped to a data symbol in $\setS$ via minimum-distance mapping, yielding the index of the detected data symbol of stream~$k$ as
\begin{align}\label{eq:l_k^star}
l_{k}^{\star} \triangleq \argmin_{l_{k} \in \{1, \ldots, L\}} |\check{u}_{k} - s_{l_{k}}|,
\end{align}
where $l_{k}$ is used throughout the paper to denote the search index for the data symbol of stream~$k$.

Under the assumption that the dither vector applied at the transmitter is known at the receiver, its effect can, in some cases, be effectively mitigated at the receiver. Specifically, the dither vector cannot be removed perfectly since it is added to the precoded signal prior to the 1-bit \acp{DAC}. However, the corresponding dither-induced bias can still be subtracted from the received signal based on different linearizations of the transmitted signal (at the output of the \acp{DAC}), which are presented in Section~\ref{sec:LQP&DD}. For simplicity, we refer to this subtraction as \textit{dither removal} in the rest of the paper.

In this work, we propose new data detection methods for systems with full-resolution and 1-bit \acp{ADC} in Sections~\ref{sec:DD1} and~\ref{sec:DD2}, respectively, considering both soft-estimation-based and \ac{ML}-based methods. For soft-estimation-based data detection, we derive new combining matrices with dither removal based on the \ac{LMMSE} criterion. For \ac{ML}-based data detection, which bypasses the soft-estimation step, we develop new methods both with and without dither removal. Table~\ref{Table:DD} summarizes these methods along with their computational complexities. The proposed methods capture different trade-offs between performance and computational complexity under distinct \ac{ADC} resolutions and operating regimes. The computational complexity scales with the system parameters $M$, $N$, and $K$. In particular, soft-estimation-based and \ac{ML}-based methods exhibit linear and exponential complexity growth with $K$, respectively. Further details on the computational complexity and properties of the proposed methods are provided in Section~\ref{sec:CCA}.

\renewcommand{\arraystretch}{1.25}
\begin{table*}[t!]
\small
\centering
\setlength{\tabcolsep}{4pt}

\begin{tabular}{|C{2cm}|L{2.3cm}|L{6.6cm}|C{2.2cm}|C{3.3cm}|}
\hline
\multicolumn{2}{|c|}{\multirow{2}{*}{\textbf{Data detection method}}}
& \centering \multirow{2}{*}{\textbf{Description}}
& \multicolumn{2}{c|}{\textbf{Computational complexity}} \\ 
\cline{4-5}
 \multicolumn{1}{|c}{} 
& \multicolumn{1}{c|}{} 
& \multicolumn{1}{c|}{} & \textbf{Per channel} & \textbf{Per data symbol vector}\\
\hline
\hline

\multirow{2}{*}[-5.5mm]{\shortstack{\textbf{Full-resolution} \\ \textbf{\acp{ADC}}}}
& \textbf{\ac{BLMMSE-DR}} (Section~\ref{sec:DD1_A}) \cellcolor{blue!10}
& Soft-estimation-based data detection using the received signal after symbol-independent dither removal $\y^{\textrm{(DR)}}$ in~\eqref{eq:y(d)} \cellcolor{blue!10}
& $\mathcal{O}(M^3+MN^2)$ \cellcolor{blue!10}
& $\mathcal{O}(KL)$ \cellcolor{blue!10} \\
\hhline{|~|-|-|-|-|}

& \textbf{\ac{ML-DR}} (Section~\ref{sec:DD1_B}) \cellcolor{red!15}
& \ac{ML}-based data detection applied to the observed signal after symbol-dependent dither removal $\y_{\x}^{\textrm{(DR)}}$ in \eqref{eq:y_x(DR)}, with search over $\nu^K$ candidates ($\nu \leq L$) \cellcolor{red!15}
& -- \cellcolor{red!15}
&  $\mathcal{O}\big((M^3+MN^2)\nu^K\big)$ with $\mathbf{\Sigma}_{{\tilde{\y}}^{\textrm{(DR)}}|{\x}}$ or $\mathcal{O}\big(MN^2\nu^K\big)$ with $\diag(\mathbf{\Sigma}_{{\tilde{\y}}^{\textrm{(DR)}}|{\x}})$ \cellcolor{red!15} \\

\hline
\hline

\multirow{2}{*}[-3.5mm]{\textbf{1-bit \acp{ADC}}}
& \textbf{\ac{D-BLMMSE-DR}} (Section~\ref{sec:DD2_A}) \cellcolor{blue!10}
& Soft-estimation-based data detection using the 1-bit quantized received signal after symbol-independent dither removal $\r^{\textrm{(DR)}}$ in~\eqref{eq:r(d)} \cellcolor{blue!10}
& $\mathcal{O}(M^3 + MN^2)$ \cellcolor{blue!10}
& $\mathcal{O}(KL)$ \cellcolor{blue!10} \\
\hhline{|~|-|-|-|-|}

& \textbf{\ac{D-ML}} (Section~\ref{sec:DD2_B}) \cellcolor{red!15}
& \ac{ML}-based data detection applied to the 1-bit quantized received signal $\r$ in~\eqref{eq:r}, with search over $\nu^K$ candidates ($\nu \leq L$) \cellcolor{red!15}
& -- \cellcolor{red!15}
& $\mathcal{O}(MN^2\nu^K)$ \cellcolor{red!15} \\
\hline

\end{tabular}

\caption{Proposed data detection methods with full-resolution and 1-bit \acp{ADC}. The soft-estimation-based and \ac{ML}-based methods are highlighted in blue and red, respectively.}
\label{Table:DD}
\end{table*}

\section{Linearization of the Transmitted and Received Signals} \label{sec:LQP&DD}

The 1-bit \acp{DAC} and \acp{ADC} introduce nonlinear distortion at the transmitter and receiver, respectively, which requires the linearization of their input-output relationships to obtain a tractable framework. In this regard, we first review the Bussgang decomposition to linearize the 1-bit quantized transmitted and received signals, along with the corresponding receiver design used for the soft-estimation-based data detection baseline considered for comparison. Then, we propose a new symbol-dependent linearization of the transmitted signal, which serves as the basis for the proposed \ac{ML}-based data detection methods presented in Sections~\ref{sec:DD1} and~\ref{sec:DD2}.

\subsection{Symbol-Independent Linearization of the Transmitted and Received Signals} \label{sec:LQP&DD_A}

In this section, we utilize the well-known Bussgang decomposition \cite{Dem21}, which allows to represent the output of a nonlinear system as a scaled version of the input plus an uncorrelated distortion term.

We linearize the transmitted signal $\x_{\textrm{q}}$ in \eqref{eq:xq} with respect to $\x_{\textrm{d}}$ (and, thus, with respect to $\u$ and $\d$) as
\begin{align}\label{eq:xq (1)}
    \x_{\textrm{q}} = \F_{\tx}\x_{\textrm{d}} + \q_{\tx},
\end{align}
where $\q_{\tx} \in \Compl^{N}$ is a zero-mean, non-Gaussian distortion vector that is uncorrelated with $\x_{\textrm{d}}$ and $\F_{\tx}\in \Compl^{N \times N}$ is the symbol-independent Bussgang gain matrix at the transmitter. To obtain $\F_{\tx}$ in closed form, we assume $\x_{\textrm{d}} \sim \setC \setN (\0_{N}, \C_{\x_{\textrm{d}}})$, with
\begin{align}
    \C_{\x_{\textrm{d}}} \triangleq \mathbb{E}_{\u,\d}[\x_{\textrm{d}}\x_{\textrm{d}}^\herm] = \W\W^\herm + \sigma^2\I_N \in \Compl^{N \times N}.
\end{align}
To this end, we temporarily consider Gaussian data symbols, i.e., $ \u \sim \mathcal{CN}(\0_{K}, \mathbf{I}_K)$. Under this assumption, $\F_{\tx}$ is diagonal with closed-form expression \cite{Li17}
\begin{align}
\F_{\tx} &\triangleq \argmin_{\F} \mathbb{E}_{\x_{\textrm{d}}}\big[ \|\x_{\textrm{q}} - \F\x_{\textrm{d}} \|_2^2\big] =\sqrt{\frac{2}{\pi}\eta_{\tx}}\diag(\C_{\x_{\textrm{d}}})^{-\frac{1}{2}}.\label{eq:F_TX}
\end{align}

\begin{figure*}[t!]
\setcounter{equation}{12}
\begin{align}\label{eq:C_xq}
    \C_{\x_{\textrm{q}}} = \frac{2}{\pi}\eta_{\tx}\Big( \arcsin\big(\diag(\C_{\x_{\textrm{d}}})^{-\frac{1}{2}}\Re[\C_{\x_{\textrm{d}}}]\diag(\C_{\x_{\textrm{d}}})^{-\frac{1}{2}}\big) +j\,\arcsin\big(\diag(\C_{\x_{\textrm{d}}})^{-\frac{1}{2}}\Im[\C_{\x_{\textrm{d}}}]\diag(\C_{\x_{\textrm{d}}})^{-\frac{1}{2}}\big)\Big)
\end{align}
\hrulefill
\setcounter{equation}{15}
\begin{align}\label{eq:Cr}
    \C_{\r} \approx \frac{2}{\pi}\eta_{\rx}\Big( \arcsin\big(\diag(\C_{\y})^{-\frac{1}{2}}\Re[\C_{\y}]\diag(\C_{y})^{-\frac{1}{2}}\big) +j\,\arcsin\big(\diag(\C_{\y})^{-\frac{1}{2}}\Im[\C_{\y}]\diag(\C_{\y})^{-\frac{1}{2}}\big)\Big)
\end{align}
\hrulefill \vspace{-2mm}
\end{figure*}

With 1-bit \acp{ADC}, we further linearize the 1-bit quantized received signal $\r$ in \eqref{eq:r}~as
\setcounter{equation}{9}
\begin{align}\label{eq:r (linearized)}
    \r = \F_{\rx}\y + \q_{\rx},
\end{align}
where $\q_{\rx} \in \Compl^{M}$ is a zero-mean, non-Gaussian distortion vector that is uncorrelated with $\y$ and $\F_{\rx} \in \mathbb{C}^{M\times M}$ is the symbol-independent Bussgang gain matrix at the receiver. Following \cite{Atz23}, $\F_{\rx}$ can be well approximated as a diagonal matrix with approximate closed-form expression
\begin{align}\label{eq:F_RX}
    \F_{\rx} &\triangleq \argmin_{\F} \mathbb{E}_{\y}\big[ \|\r - \F\y \|_2^2\big] \approx \sqrt{\frac{2}{\pi}\eta_{\rx}}\diag(\C_{\y})^{-\frac{1}{2}},
\end{align}
with
\begin{align}\label{eq:Cy}
    \C_{\y} \triangleq \mathbb{E}[\y\y^\herm] = \rho\H\C_{\x_{\textrm{q}}}\H^\herm + \I_M \in \mathbb{C}^{M\times M}
\end{align}
and where $\C_{\x_{\textrm{q}}} \triangleq \mathbb{E}[\x_{\textrm{q}}\x_{\textrm{q}}^\herm] \in \mathbb{C}^{N\times N}$ is shown in \eqref{eq:C_xq} at the top of the page.

Based on the linearizations in \eqref{eq:xq (1)} and \eqref{eq:r (linearized)}, we obtain the \ac{BLMMSE} and \ac{D-BLMMSE} receivers, which minimize the \ac{MSE} between $\check{\u}$ in \eqref{eq:hat_s} and $\u$ for full-resolution and 1-bit \acp{ADC}, respectively. The \ac{BLMMSE} combining matrix applied to $\y$ in \eqref{eq:y} is given by
\setcounter{equation}{13}
\begin{align}
    \V_{\textrm{BLMMSE}} = \sqrt{\rho} \C_{\y}^{-1}\H\F_{\tx}\W,
\end{align}
with $\F_{\tx}$ and $\C_{\y}$ given in \eqref{eq:F_TX} and \eqref{eq:Cy}, respectively. Similarly, the \ac{D-BLMMSE} combining matrix applied to $\r$ in \eqref{eq:r} is given by \cite{Atz23}
\begin{align}
    \V_{\textrm{D-BLMMSE}} = \sqrt{\rho} \C_{\r}^{-1}\F_{\rx}\H\F_{\tx}\W,
\end{align}
with $\F_{\rx}$ given in \eqref{eq:F_RX} and where $\C_{\r} \triangleq \mathbb{E}[\r\r^\herm] \in \mathbb{C}^{M\times M}$ can be approximated as shown in \eqref{eq:Cr} at the top of the page.\footnote{As observed in \cite{Atz23}, the received signal $\y$ can be accurately approximated as Gaussian for $N\geq 8$, which justifies the approximation in \eqref{eq:Cr}.}

\begin{figure*}[t!]
\setcounter{equation}{18}
\begin{align}\label{eq:Cxdxq|x (1)}
& [\C_{\x_{\textrm{d}}\x_{\textrm{q}}|\x}]_{n,m} \nonumber \\
& = [\C_{\x_{\textrm{d}}\x_{\textrm{q}}|\x}^{( \Re, \Re)}]_{n,m} + [\C_{\x_{\textrm{d}}\x_{\textrm{q}}|\x}^{( \Im, \Im)}]_{n,m} - j\,[\C_{\x_{\textrm{d}}\x_{\textrm{q}}|\x}^{( \Re, \Im)}]_{n,m} + j\,[\C_{\x_{\textrm{d}}\x_{\textrm{q}}|\x}^{( \Im, \Re)}]_{n,m} \\
& = \begin{cases}
    \sqrt{\frac{\eta}{2}} \Big( \sqrt{\frac{\sigma^2}{\pi}} \Big(\exp \big( - \big(\frac{\Re[x_m]}{\sigma}\big)^2 \big) + \exp \big( - \big(\frac{\Im[x_m]}{\sigma} \big)^2 \big)\Big) + \Re \big[x_n\erf_{\textrm{c}} \big(\frac{x_m^*}{\sigma} \big) \big] + j \, \Im \big[x_n\erf_{\textrm{c}} \big(\frac{x_m^*}{\sigma} \big) \big] \Big) & \ \textrm{if}~m=n, \\
  \sqrt{\frac{\eta}{2}} \big(\Re \big[x_n\erf_{\textrm{c}} \big(\frac{x_m^*}{\sigma} \big) \big] +j \, \Im \big[x_n\erf_{\textrm{c}} \big(\frac{x_m^*}{\sigma} \big) \big] \big) & \ \textrm{otherwise},
    \end{cases}\label{eq:Cxdxq|x (2)}
\end{align}
\hrulefill \vspace{-2mm}
\setcounter{equation}{26}
    \begin{align}
   \nonumber & \C_{\tilde{\y}|\x} = \\
   & \hspace{-1mm} \begin{bmatrix}
    \sqrt{\rho}\f^{(\Re)}(\sqrt{\rho}\f^{(\Re)} \! + \! \mub_{\n}^{(\Re)})^\tran \! + \! \sqrt{\rho}\mub_{\n}^{(\Re)}(\f^{(\Re)})^\tran \! + \! \C_{\tilde{\n}}^{(\Re,\Re)} \!\!\! & \sqrt{\rho}\f^{(\Re)}( \sqrt{\rho}\f^{(\Im)} \! + \! \mub_{\n}^{(\Im)})^\tran \! + \! \sqrt{\rho}\mub_{\n}^{(\Re)}(\f^{(\Im)})^\tran \! + \! \C_{\tilde{\n}}^{(\Re,\Im)} \\
    \sqrt{\rho}( \sqrt{\rho}\f^{(\Im)} \! + \! \mub_{\n}^{(\Im)})(\f^{(\Re)})^\tran \! + \! \sqrt{\rho}\f^{(\Im)}(\mub_{\n}^{(\Re)})^\tran \! + \! (\C_{\tilde{\n}}^{(\Re,\Im)})^\tran \!\!\! & \sqrt{\rho}\f^{(\Im)}( \sqrt{\rho}\f^{(\Im)} \! + \! \mub_{\n}^{(\Im)})^\tran \! + \! \sqrt{\rho}\mub_{\n}^{(\Im)}(\f^{(\Im)})^\tran \! + \! \C_{\tilde{\n}}^{(\Im,\Im)}
    \end{bmatrix}\label{eq:Cy'|x}
\end{align}
\hrulefill \vspace{-2mm}
\end{figure*}

\subsection{Symbol-Dependent Linearization of the Transmitted Signal}\label{sec:LQP&DD_B}

Here, we assume that the data symbol vector $\u$ (and, consequently, $\mathbf{x}$) is fixed, which implies that the dithered precoded signal conditioned on $\mathbf{x}$ satisfies $\mathbf{x}_{\textrm{d}}|\x \sim \mathcal{CN}(\mathbf{x}, \sigma^2 \mathbf{I}_N)$. We linearize the transmitted signal $\mathbf{x}_{\textrm{q}}$ in \eqref{eq:xq} for a given $\mathbf{x}$ as
\setcounter{equation}{16}
\begin{align}\label{eq:xq (2)}
     \x_{\textrm{q}} = \G_{\tx}(\x)\x_{\textrm{d}} + \p_{\tx},
\end{align}
where $\p_{\tx} \in \mathbb C^{N}$ is a non-Gaussian distortion vector that is uncorrelated with $\x_{\textrm{d}}$. Furthermore, $\G_{\tx}(\x)\in \mathbb{C}^{N\times N}$ is the symbol-dependent Bussgang gain matrix at the transmitter obtained as
\begin{align}
\hspace{-2mm} \G_{\tx}(\x)& \triangleq \argmin_{\G} \mathbb{E}_{\d}\big[ \|\x_{\textrm{q}} \! - \! \G\x_{\textrm{d}} \|_2^2 | \x \big] = \C_{\x_{\textrm{d}}\x_{\textrm{q}}|\x}^\herm \C_{\x_{\textrm{d}}|\x}^{-1}\label{eq:G(x)},
\end{align}
where $\C_{\x_{\textrm{d}}|\x} \triangleq \mathbb E_{\d}[\x_{\textrm{d}}\x_{\textrm{d}}^\herm|\x] = \x\x^\herm + \sigma^2\I_N \in \mathbb{C}^{N\times N}$ is the symbol-dependent auto-correlation matrix of $\x_{\textrm{d}}$ and $\C_{\x_{\textrm{d}}\x_{\textrm{q}}|\x} \triangleq \mathbb{E}_{\d}[\x_{\textrm{d}}\x_{\textrm{q}}^\herm|\x] \in \mathbb{C}^{N\times N}$ is the symbol-dependent cross-correlation matrix between $\x_{\textrm{d}}$ and $\x_{\textrm{q}}$; the latter is detailed in \eqref{eq:Cxdxq|x (1)}--\eqref{eq:Cxdxq|x (2)} at the top of the next page, with $\C_{\x_{\textrm{d}}\x_{\textrm{q}}|\x}^{(\mathrm{A}, \mathrm{B})} \triangleq \mathbb{E}_{\d}\big[\mathrm{A}[\x_{\textrm{d}}]\mathrm{B}[\x_{\textrm{q}}]^\tran|\x\big] \in \mathbb{R}^{N\times N}$ derived in Appendix~\ref{sec:App A}.

\section{Data Detection with Full-Resolution \acp{ADC}}\label{sec:DD1}

In this section, we consider a system model with 1-bit \acp{DAC} and full-resolution \acp{ADC}. Building on the symbol-independent linearization described in Section~\ref{sec:LQP&DD_A}, we develop a soft-estimation-based method with symbol-independent dither removal in Section~\ref{sec:DD1_A}. Then, building on the symbol-dependent linearization described in Section~\ref{sec:LQP&DD_B}, we develop an \ac{ML}-based method in Section~\ref{sec:DD1_B}, which directly detects the data symbol vector $\u$ from the received signal $\y$ in \eqref{eq:y}. We begin by presenting this method without dither removal, deriving the full likelihood function of the received signal along with its statistics: these results are later used in Section~\ref{sec:DD2_B} and also provide the basis for the subsequent methods. Next, we incorporate symbol-dependent dither removal, which also yields a simplified formulation. Lastly, we introduce a low-complexity variant that searches over a subset of candidates identified by the method in Section~\ref{sec:DD1_A}.

\subsection{\ac{BLMMSE} with Dither Removal (\ac{BLMMSE-DR})} \label{sec:DD1_A}

Here, we assume that the dither vector applied at the transmitter is known at the receiver, so that dither removal can be performed.\footnote{In practice, the receiver only needs to know the seed of the corresponding random vector, which can be agreed in advance or shared through a low-rate feedback channel between the transmitter and receiver.} Based on the symbol-independent linearization of the transmitted signal in \eqref{eq:xq (1)}, we subtract the term associated with the dither vector $\d$ from the received signal $\y$ in \eqref{eq:y}. Let $\y^{\textrm{(DR)}}$ denote the received signal after symbol-independent dither removal, defined as
\setcounter{equation}{20}
\begin{align}\label{eq:y(d)}
    \y^{\textrm{(DR)}} \triangleq \y - \sqrt{\rho}\H\F_{\tx}\d \in \Compl^{M},
\end{align}
with $\F_{\tx}$ defined in \eqref{eq:F_TX}. In the data detection via \textit{BLMMSE with dither removal (BLMMSE-DR)}, a soft estimate of the data symbol vector is obtained as $\check{\u}_{\textrm{BLMMSE-DR}} = \V_{\textrm{BLMMSE-DR}}^\herm\y^{\textrm{(DR)}}$, where the combining matrix is determined as
\begin{align}
\label{eq:V design (1)}
    \V_{\textrm{BLMMSE-DR}} &\triangleq\argmin_{\V} \mathbb{E}_{\u,\z, \d}\big[\|\V^\herm\y^{\textrm{(DR)}} - \u \|_2^2 \big] \\ \label{eq:V_DBLMMSE}
    &= \sqrt{\rho} (\C_{\y} - \rho\sigma^2\H\F_{\tx}^2\H^\herm)^{-1}\H\F_{\tx}\W.
\end{align}
Finally, the indices of the detected data symbols are extracted as in \eqref{eq:l_k^star}.

\subsection{\ac{ML} with Dither Removal (\ac{ML-DR})} \label{sec:DD1_B}

Here, we switch our attention to \ac{ML}-based data detection. First, we assume the dither vector is not known at the receiver and derive the full likelihood function of the received signal $\y$ in \eqref{eq:y} along with its statistics, which are later used in Section~\ref{sec:DD2_B}. Then, we enhance this method by incorporating symbol-dependent dither removal, which also yields a simplified formulation. Lastly, we introduce a low-complexity variant with reduced search space.

\smallskip

\textbf{\textit{\ac{ML}.}} Based on the symbol-dependent linearization of the transmitted signal in \eqref{eq:xq (2)} described in Section~\ref{sec:LQP&DD_B}, we rewrite $\y$ in \eqref{eq:y} for a given $\x$ as
\setcounter{equation}{23}
\begin{align}\label{eq:y_decomposed}
    \y = \sqrt{\rho}\H \G_{\tx}(\x)\x + \n,
\end{align}
with $\G_{\tx}(\x)$ defined in \eqref{eq:G(x)} and where $\n \triangleq \sqrt{\rho}\H \big(\G_{\tx}(\x)\d + \p_{\tx}\big) + \z \in \mathbb{C}^{M}$ is the effective noise comprising dither, quantization distortion, and \ac{AWGN}. For large $N$ and sufficiently high dither power $\sigma^2$, the elements of $\p_{\tx}$ become weakly correlated across antennas \cite{Sax19}, which makes $\n$ approximately Gaussian. While this approximation is only asymptotically exact for $N \to \infty$, our simulations indicate that it is quite accurate even for moderate dimensions, specifically for $N\geq 16$ and $\sigma^2 \geq-10$~dBm. Hence, we define $\tilde{\n} \triangleq \big[\Re[\n]^\tran,\Im[\n]^\tran\big]^\tran \in \mathbb{R}^{2M}$, which approximately follows $\tilde{\n} \sim \mathcal{N} \big(\boldsymbol{\mu}_{\tilde{\n}}(\x),\mathbf{\Sigma}_{\tilde{\n}}(\x)\big)$, with mean $\boldsymbol{\mu}_{\tilde{\n}}(\x) \in \mathbb{R}^{2M}$ and covariance matrix $\mathbf{\Sigma}_{\tilde{\n}}(\x) \in \mathbb{R}^{2M \times 2M}$ derived in \eqref{eq:mu_n'} and \eqref{eq:Sigma_n'}, respectively, in Appendix~\ref{sec:App B}.

Further define $\tilde{\y} \triangleq \big[\Re[\y]^\tran, \Im[\y]^\tran\big]^\tran \in \mathbb{R}^{2M}$, $\f^{(\mathrm{A})} \triangleq \mathrm{A} \big[\H\G_{\tx}(\x)\x \big] \in \mathbb{R}^{M}$, along with $\mub_{\n}^{(\mathrm{A})} \triangleq \mathbb{E}\big[\mathrm{A}[\n]\big] \in \mathbb{R}^{M}$ and $\C_{\tilde{\n}}^{(\mathrm{A,B})} \triangleq \mathbb{E}\big[\mathrm{A}[\n]\mathrm{B}[\n]^\tran\big] \in \mathbb{R}^{M\times M}$, which are derived in \eqref{eq:E[Re[zt]]}--\eqref{eq:E[Im[zt]]} and \eqref{eq:Cnt_Re}, respectively, in Appendix~\ref{sec:App B}. From \eqref{eq:y_decomposed}, we have $\tilde{\y}|\x \sim \mathcal{N}(\boldsymbol{\mu}_{\tilde{\y}|\x},\mathbf{\Sigma}_{\tilde{\y}|\x})$, with mean
\begin{align}
    \boldsymbol{\mu}_{\tilde{\y}|\x} &\triangleq \mathbb{E}[\tilde{\y}|\x] = \begin{bmatrix}
    \sqrt{\rho} \f^{(\Re)} + \mub_{\n}^{(\Re)}\\
      \sqrt{\rho} \f^{(\Im)} + \mub_{\n}^{(\Im)}
    \end{bmatrix} \in \mathbb{R}^{2M} \label{eq:mu}
\end{align}
and covariance matrix
\begin{align} \label{eq:Sigma_y'}
    \mathbf{\Sigma}_{\tilde{\y}|\x} \triangleq \C_{\tilde{\y}|\x} - \boldsymbol{\mu}_{\tilde{\y}|\x}\boldsymbol{\mu}_{\tilde{\y}|\x}^\tran \in \mathbb{R}^{2M\times 2M},
\end{align}
where $\C_{\tilde{\y}|\x} \triangleq \mathbb{E}[\tilde{\y}\tilde{\y}^\tran|\x] \in \mathbb{R}^{2M\times 2M}$ is detailed in \eqref{eq:Cy'|x} at the top of the page. Finally, recalling that the precoded signal is expressed as $\x = \W\u$, the \ac{ML} data detection problem is formulated as
\setcounter{equation}{27}
\begin{align}\label{eq:hat_s_ML}
    \hat{\u}_{\textrm{ML}} \triangleq \argmin_{\u \in \setS^{K}} \, \zeta_{\textrm{ML}}(\x),
\end{align}
where
\begin{align}
    \hspace{-2mm} \zeta_{\textrm{ML}}(\x) \triangleq ( \tilde{\y} - \boldsymbol{\mu}_{\tilde{\y}|\x})^\tran \mathbf{\Sigma}_{\tilde{\y}|\x}^{-1}( \tilde{\y} -  \boldsymbol{\mu}_{\tilde{\y}|\x}) + \logdet (\mathbf{\Sigma}_{\tilde{\y}|\x}) \label{eq:zeta(x)}
\end{align}
is the objective function obtained from the negative log-likelihood of $\tilde{\y}|\x$. The total computational complexity (per channel and per data symbol vector) of this method is $\mathcal{O}\big((M^3+MN^2)L^K\big)$.

\smallskip

\textbf{\textit{\ac{ML-DR}.}} As for \ac{BLMMSE-DR}, we assume that the dither vector applied at the transmitter is known at the receiver. Based on the symbol-dependent linearization of the transmitted signal in \eqref{eq:xq (2)}, we subtract the term associated with the dither vector $\d$ from the received signal $\y$ in \eqref{eq:y}. In this regard, for a given $\x$, let $\y_{\x}^{\textrm{(DR)}}$ denote the observed signal after symbol-dependent dither removal from $\y$, defined as
\begin{align}
    \y_{\x}^{\textrm{(DR)}} \triangleq \y - \sqrt{\rho}\H\G_{\tx}(\x)\d \in \Compl^{M},\label{eq:y_x(DR)}
\end{align}
with $\G_{\tx}(\x)$ defined in \eqref{eq:G(x)}. Now, defining ${\tilde{\y}}_{\x}^{\textrm{(DR)}} \triangleq \big[\Re[\y_{\x}^{\textrm{(DR)}}]^\tran, \Im[\y_{\x}^{\textrm{(DR)}}]^\tran\big]^\tran \in \mathbb{R}^{2M}$, we have ${\tilde{\y}}_{\x}^{\textrm{(DR)}} \sim \mathcal{N}(\boldsymbol{\mu}_{{\tilde{\y}}_{\x}^{\textrm{(DR)}}},\mathbf{\Sigma}_{{\tilde{\y}}^{\textrm{(DR)}}|{\x}})$, with mean $\boldsymbol{\mu}_{{\tilde{\y}}_{\x}^{\textrm{(DR)}}}\triangleq\mathbb{E}[{\tilde{\y}}_{\x}^{\textrm{(DR)}}] = \boldsymbol{\mu}_{\tilde{\y}|\x}$ (see \eqref{eq:mu}) and covariance matrix  $\mathbf{\Sigma}_{{\tilde{\y}}^{\textrm{(DR)}}|{\x}} \triangleq \C_{{\tilde{\y}}_{\x}^{\textrm{(DR)}}} - \boldsymbol{\mu}_{{\tilde{\y}}_{\x}^{\textrm{(DR)}}} \boldsymbol{\mu}_{{\tilde{\y}}_{\x}^{\textrm{(DR)}}}^\tran \in \mathbb{R}^{2M\times 2M}$. For the latter, $\C_{{\tilde{\y}}_{\x}^{\textrm{(DR)}}} \triangleq \mathbb{E}[{\tilde{\y}}_{\x}^{\textrm{(DR)}}({\tilde{\y}}_{\x}^{\textrm{(DR)}})^\tran] \in \mathbb{R}^{2M\times 2M}$ can be derived in a form analogous to \eqref{eq:Cy'|x}. In this case, the symbol-dependent dither removal allows to simplify $\C_{\tilde{\n}}^{(\mathrm{A},\mathrm{B})}$ from its full expression in \eqref{eq:Cnt_Re} to
\begin{align}
    \C_{\tilde{\n}}^{(\mathrm{A},\mathrm{B})} = \rho\mathbb{E}\big[\mathrm{A}[\H\p_{\tx}]\mathrm{B}[\H\p_{\tx}]^\tran\big] + \frac{\delta_{\mathrm{A},\mathrm{B}}}{2}\I_M,\label{eq:C_tilde_n}
\end{align}
with $\mathbb{E}\big[\mathrm{A}[\p_{\tx}]\mathrm{B}[\p_{\tx}]^\tran\big]$ derived in \eqref{eq:Cqx_re} in Appendix~\ref{sec:App B}. Then, the \textit{ML with dither removal (\ac{ML-DR})} data detection problem is formulated as
\begin{align}\label{eq:ML-DR}
    \hat{\u}_{\textrm{ML-DR}} \triangleq \argmin_{\u \in \setS^K} \, \zeta_{\textrm{ML-DR}}(\x),
\end{align}
where
\begin{align}
    \nonumber \zeta_{\textrm{ML-DR}}(\x) & \triangleq ({\tilde{\y}}_{\x}^{\textrm{(DR)}} - \boldsymbol{\mu}_{{\tilde{\y}}_{\x}^{\textrm{(DR)}}})^\tran\mathbf{\Sigma}_{{\tilde{\y}}^{\textrm{(DR)}}|{\x}}^{-1}( {\tilde{\y}}_{\x}^{\textrm{(DR)}} -  \boldsymbol{\mu}_{{\tilde{\y}}_{\x}^{\textrm{(DR)}}}) \\
    & \phantom{=} \ + \logdet \big(\mathbf{\Sigma}_{{\tilde{\y}}^{\textrm{(DR)}}|{\x}}\big)\label{eq:zeta_MLDR}
\end{align}
is the objective function obtained from the negative log-likelihood of $\tilde{\y}_{\x}^{\textrm{(DR)}}$.

Interestingly, we observe from our numerical results in Section~\ref{sec:NR} that the off-diagonal elements of $\mathbf{\Sigma}_{{\tilde{\y}}^{\textrm{(DR)}}|{\x}}$ in \eqref{eq:ML-DR} can be neglected without any noticeable performance loss. Specifically, although these elements are not strictly negligible, the dither removal step increases the separation between likelihood values corresponding to different values of $\x$. Consequently, even the diagonal approximation $\diag(\mathbf{\Sigma}_{{\tilde{\y}}^{\textrm{(DR)}}|{\x}})$ preserves the relative ordering of candidate likelihoods, maintaining nearly the same performance. This simplification mitigates the computational complexity per data symbol vector by avoiding matrix inversion and log-determinant operations, decreasing the total computational complexity (per channel and per data symbol vector) from $\mathcal{O}\big((M^3+MN^2)L^K\big)$ to $\mathcal{O}(MN^2L^K)$.

\smallskip

\textbf{\textit{\ac{ML-DR} with reduced search space.}} Both the \ac{ML} and \ac{ML-DR} methods described above perform an exhaustive search over all the $L^{K}$ possible data symbol vectors $\u$ and thus their computational complexity grows exponentially with $K$. This computational burden can be mitigated by restricting the search to a subset of candidates around a suitable predefined soft-estimated symbol, for example, identified using \ac{BLMMSE-DR} (see Section~\ref{sec:DD1_A}).

Recalling the data detection via \ac{BLMMSE-DR} in Section~\ref{sec:DD1_A}, we consider the $\nu \leq L$ values of $s_{l_{k}}$ that lie closest to stream~$k$'s soft-estimated symbol $\check{u}_{\textrm{BLMMSE-DR},k}$. From \eqref{eq:l_k^star}, let $l_{k}^{(i)}$ denote the index of the candidate data symbol of stream~$k$ with the $i$th smallest distance $|\check{u}_{\textrm{BLMMSE-DR},k} - s_{l_{k}^{(i)}}|$, with $i \in \{1, \ldots, \nu\}$, and define $\setS'_{k} = \{s_{l_{k}^{(i)}}\}_{i=1}^{\nu}$.  Then, the data detection problem in \eqref{eq:ML-DR} simplifies to
\begin{align}\label{eq:ML-DR (2)}
    \hat{\u}_{\textrm{ML-DR}} = \argmin_{\u \in \prod_{k=1}^{K} \setS'_{k}} \, \zeta_{\textrm{ML-DR}}(\x).
\end{align}
As a result of the reduced search space, the total computational complexity decreases from $\mathcal{O}(MN^2L^K)$ to $\mathcal{O}(MN^2\nu^K)$. In our numerical results in Section~\ref{sec:NR}, we consider \ac{ML-DR} with reduced search space specifying $\nu \in \{1, \ldots, L\}$, where the extreme cases $\nu=1$ and $\nu=L$ correspond to \ac{BLMMSE-DR} and full-search \ac{ML-DR}, respectively.

\section{Data Detection with 1-Bit \acp{ADC}}\label{sec:DD2}

In this section, we consider a doubly 1-bit quantized system model with 1-bit \acp{DAC} and \acp{ADC}, adapting the data detection methods proposed in Section~\ref{sec:DD1} for full-resolution \acp{ADC}. Building on the symbol-independent linearization described in Section~\ref{sec:LQP&DD_A}, we develop a soft-estimation-based method with symbol-independent dither removal in Section~\ref{sec:DD2_A}. Then, we develop an \ac{ML}-based method in Section~\ref{sec:DD2_B}, which directly detects the data symbol vector $\u$ from the 1-bit quantized received signal $\r$ in \eqref{eq:r}. Lastly, we introduce a low-complexity variant that searches over a subset of candidates identified by the method in Section~\ref{sec:DD2_A}.

\subsection{Doubly 1-Bit Quantized \ac{BLMMSE} with Dither Removal (\ac{D-BLMMSE-DR})} \label{sec:DD2_A}

Similar to \ac{BLMMSE-DR} presented in Section~\ref{sec:DD1_A}, we assume the dither vector applied at the transmitter is known at the receiver. Based on the symbol-independent linearizations of the transmitted and received signals in \eqref{eq:xq (1)} and \eqref{eq:r (linearized)}, respectively, we subtract the term associated with the dither vector $\d$ from the 1-bit quantized received signal $\r$ in \eqref{eq:r}. Let $\r^{\textrm{(DR)}}$ denote the 1-bit quantized received signal after symbol-independent dither removal, defined as
\begin{align}\label{eq:r(d)}
    \r^{\textrm{(DR)}} \triangleq \r - \sqrt{\rho}\F_{\rx}\H\F_{\tx}\d \in \Compl^{M},
\end{align}
with $\F_{\tx}$ and $\F_{\rx}$ defined in \eqref{eq:F_TX} and \eqref{eq:F_RX}, respectively. In the data detection via \textit{doubly 1-bit quantized BLMMSE with dither removal (D-BLMMSE-DR)}, a soft estimate of the data symbol vector is obtained as $\check{\u}_{\textrm{D-BLMMSE-DR}} = \V_{\textrm{D-BLMMSE-DR}}^\herm\r^{\textrm{(DR)}}$, where the combining matrix is determined as
\begin{align}
\label{eq:V design (2)}
    \V_{\textrm{D-BLMMSE-DR}} &\triangleq \argmin_{\V} \mathbb{E}_{\u,\z, \d}\big[\|\V^\herm\r^{\textrm{(DR)}} - \u \|_2^2 \big] \\
    &=\nonumber\sqrt{\rho} (\C_{\r}-\rho\sigma^2\F_{\rx}\H\F_{\tx}^2\H^\herm\F_{\rx})^{-1}\\ & \phantom{=} \ \times\F_{\rx}\H\F_{\tx}\W. \label{eq:V_DBLMMSE_doubly}
\end{align}
Finally, the indices of the detected data symbols are extracted as in \eqref{eq:l_k^star}. We note that the dither removal step in \eqref{eq:r(d)} and the resulting combining matrix in \eqref{eq:V_DBLMMSE_doubly} are equivalent to their counterparts for full-resolution \acp{ADC} in \eqref{eq:y(d)} and \eqref{eq:V_DBLMMSE}, respectively, the only differences being the observed signal (i.e., $\y$ in \eqref{eq:y} or $\r$ in \eqref{eq:r}) and the presence of the symbol-independent Bussgang gain matrix at the receiver in \eqref{eq:F_RX}.

\subsection{Doubly 1-Bit Quantized \ac{ML} (\ac{D-ML})} \label{sec:DD2_B}

Here, we turn to \ac{ML}-based data detection. We begin by noting that the symbol-dependent dither removal adopted in Section~\ref{sec:DD1_B} for full-resolution \acp{ADC} does not extend to the case of 1-bit \acp{ADC}. The reason is that the additive-noise model in \eqref{eq:y_decomposed}, which applies to the received signal $\y$, does not hold for the 1-bit quantized received signal $\r$. As a result, the quantization distortion introduced by the 1-bit \acp{ADC} remains input-dependent for any dither distribution \cite{Rap19}, and the term associated with the dither vector $\d$ does not appear as an independent additive contribution that can be removed by direct subtraction. Accordingly, we first use the statistics of the received signal $\y$ (without dither removal) given in Section~\ref{sec:DD1_B} to develop an approximated \ac{ML}-based method based on the 1-bit quantized received signal $\r$ in \eqref{eq:r}. Then, we introduce a low-complexity variant with reduced search space.

\textbf{\textit{\ac{D-ML}.}} Defining $\tilde{\r} \triangleq \big[\Re[\r]^\tran,\Im[\r]^\tran\big]^\tran \in \mathbb{R}^{2M}$ and recalling the definition of $\tilde{\y}$ from Section~\ref{sec:DD1_B}, we have $\tilde{\r} = \sqrt{\frac{\eta_{\rx}}{2}}\sgn(\tilde{\y})$, whereby each element of $\tilde{\r}$ depends only on the sign of the corresponding element of $\tilde{\y}$. Moreover, let $\mathcal{P} \triangleq \big\{m \in \{1,\dots,2M\}: \tilde{y}_m \ge 0\big\}$ and $\mathcal{\bar{P}} \triangleq \big\{m \in \{1,\dots,2M\}: \tilde{y}_{m} < 0\big\}$ denote, respectively, the sets of the indices corresponding to the positive and negative elements of $\tilde{\y}$, with $|\mathcal{P}| + |\mathcal{\bar{P}}| = 2M$, and where $\tilde{y}_m$ is the $m$th element of $\tilde{\y}$. Since observing the elements of $\tilde{\r}$ is equivalent to observing the sign of the elements of $\tilde{\y}$, the likelihood function of the 1-bit quantized received signal $\tilde{\r}$ for a given $\x$ can be defined~as
\begin{align}\label{eq:L(x)}
\zeta_{\textrm{D-ML}}(\x)
&\triangleq \Pr[
\tilde{y}_m \ge 0, \forall m\in\mathcal{P}, \ 
\tilde{y}_{m} < 0, \forall m\in\mathcal{\bar{P}} | \x].
\end{align}
Now, we focus on the statistics of $\tilde{\y}$ to derive the probability in \eqref{eq:L(x)}. From Section~\ref{sec:DD1_B}, we rewrite $\tilde{\y}$ in  \eqref{eq:y_decomposed} as
\begin{align}\label{eq:y_tilde}
    \tilde{\y} = \boldsymbol{\mu}_{\tilde{\y}|\x} + \tilde{\n}',
\end{align}
with $\mub_{\tilde{\y}|\x}$ given in \eqref{eq:mu} and $\tilde{\n}' \triangleq \tilde{\n} - \boldsymbol{\mu}_{\tilde{\n}}(\x) \in \mathbb{R}^{2M}$,  with $\boldsymbol{\mu}_{\tilde{\n}}(\x)$ derived in \eqref{eq:mu_n'} in Appendix~\ref{sec:App B}. Since $\tilde{\n}$ approximately follows $\mathcal{N} \big(\boldsymbol{\mu}_{\tilde{\n}}(\x),\mathbf{\Sigma}_{\tilde{\n}}(\x)\big)$ (see Section~\ref{sec:DD1_B}), we have that $\tilde{\n}'$ approximately follows $\mathcal{N} \big(\0_{2M},\mathbf{\Sigma}_{\tilde{\n}}(\x)\big)$, with $\mathbf{\Sigma}_{\tilde{\n}}(\x)$ derived in \eqref{eq:Sigma_n'} in Appendix~\ref{sec:App B}.

To make \eqref{eq:L(x)} tractable, we derive a closed-form approximation by neglecting the off-diagonal elements of $\mathbf{\Sigma}_{\tilde{\n}}(\x)$ and treating the elements of $\tilde{\n}'$ as independent random variables, which in turn implies independent elements of $\tilde{\y}$.\footnote{Here, we neglect the off-diagonal elements of $\mathbf{\Sigma}_{\tilde{\n}}(\x)$ for mathematical tractability, whereas in Section~\ref{sec:DD1_B} we neglect the off-diagonal elements of $\mathbf{\Sigma}_{{\tilde{\y}}^{\textrm{(DR)}}|{\x}}$ to reduce the computational complexity per data symbol vector.} Although these off-diagonal elements are not strictly negligible, this approximation yields a tractable likelihood expression that works well for large $N$ and sufficiently large $\sigma^2$, where the effective-noise components become weakly correlated. Let $\tilde{r}_m$, $\tilde{n}'_m$, and $\mu_{\tilde{\y}|\x,m}$ denote the $m$th elements of $\tilde{\r}$, $\tilde{\n}'$, and $\boldsymbol{\mu}_{\tilde{\y}|\x}$, respectively, and let $\mu_{\tilde{\y}|\x,m}^{\textrm{(sr)}} \triangleq \sqrt{\frac{2}{\eta_{\rx}}}\tilde{r}_m \mu_{\tilde{\y}|\x,m}$ denote the sign-refined version of $\mu_{\tilde{\y}|\x,m}$. Then, we approximate $\zeta_{\textrm{D-ML}}(\x)$ in \eqref{eq:L(x)} as 
\begin{align}
\zeta_{\textrm{D-ML}}(\x) & \approx\nonumber\Pr[\mu_{\tilde{\y}|\x,m} + \tilde{n}'_m \ge 0, \forall m\in\mathcal{P}] \\& \phantom{=} \ \times \Pr[\mu_{\tilde{\y}|\x,m} + \tilde{n}'_{m} < 0, \forall m\in\mathcal{\bar{P}}]\label{eq:L(x) a0} \\&=
\nonumber\Pr[\mu_{\tilde{\y}|\x,m}^{\textrm{(sr)}} \geq -\tilde{n}'_m, \forall m\in\mathcal{P}] \\& \phantom{=} \ \times
\Pr[\mu_{\tilde{\y}|\x,m}^{\textrm{(sr)}} \geq \tilde{n}'_{m}, \forall m\in\mathcal{\bar{P}}] \label{eq:L(x) a} \\& =
\nonumber\Pr[\mu_{\tilde{\y}|\x,m}^{\textrm{(sr)}} \geq -\tilde{n}'_m, \forall m\in\mathcal{P}] \\& \phantom{=} \ \times
\Pr[\mu_{\tilde{\y}|\x,m}^{\textrm{(sr)}} \geq -\tilde{n}'_{m}, \forall m\in\mathcal{\bar{P}}]\label{eq:L(x) b} \\&=
\prod_{m=1}^{2M}\Phi(\bar{\mu}_{\tilde{\y}|\x,m}) \label{eq:L(x) c},
\end{align}
where $\bar{\mu}_{\tilde{\y}|\x,m}$ denotes the $m$th element of $\bar{\mub}_{\tilde{\y}|\x} \triangleq \diag\big(\mathbf{\Sigma}_{\tilde{\n}}(\x)\big)^{-\frac{1}{2}}\mub_{\tilde{\y}|\x}^{\textrm{(sr)}}$. In the above derivations: the approximation in \eqref{eq:L(x) a0} follows from neglecting the off-diagonal elements of $\mathbf{\Sigma}_{\tilde{\n}}(\x)$; \eqref{eq:L(x) a} follows from the definition of $\mu_{\tilde{\y}|\x,m}^{\textrm{(sr)}}$; \eqref{eq:L(x) b} is due to the fact that the random variables $\tilde{n}'_m$ and $-\tilde{n}'_m$ have the same distribution, which yields $\Pr(\mu_{\tilde{\y}|\x,m}^{\textrm{(sr)}}\geq \tilde{n}'_m) = \Pr(\mu_{\tilde{\y}|\x,m}^{\textrm{(sr)}}\geq -\tilde{n}'_m)$; and \eqref{eq:L(x) c} follows from the fact that $\{\tilde{n}'_m\}_{m=1}^{2M}$ are zero-mean independent random variables with variance given by the corresponding diagonal elements of $\mathbf{\Sigma}_{\tilde{\n}}(\x)$. Finally, recalling that the precoded signal is expressed as $\x = \W\u$, the \textit{doubly 1-bit quantized ML (D-ML)} data detection problem is formulated as
\begin{align}\label{eq:Doubly ML}
     \hat{\u}_{\textrm{D-ML}} &\triangleq \argmax_{\u \in \setS^{K}} \underbrace{\sum_{m=1}^{2M} \mathrm{log} \big(\Phi(\bar{\mu}_{\tilde{\y}|\x,m})\big)}_{\approx \mathrm{log} (\zeta_{\textrm{D-ML}}(\x))}.
\end{align}
The total computational complexity (per channel and per data symbol vector) of this method is $\mathcal{O}(MN^2L^K)$.

\smallskip

\textbf{\textit{\ac{D-ML} with reduced search space.}} Similar to \ac{ML-DR} in Section~\ref{sec:DD1_B}, we can mitigate the exponential complexity growth with $K$ by restricting the search to a subset of candidates around a suitable predefined soft-estimated symbol, for example, identified using \ac{D-BLMMSE-DR} (see Section~\ref{sec:DD2_A}).

Recalling the data detection via \ac{D-BLMMSE-DR} in Section~\ref{sec:DD2_A}, we consider the $\nu \leq L$ values of $s_{l_{k}}$ that lie closest to stream~$k$'s soft-estimated symbol $\check{u}_{\textrm{D-BLMMSE-DR},k}$. Furthermore, we define $l_{\textrm{D},k}^{(i)}$ similar to $l_{k}^{(i)}$ in Section~\ref{sec:DD1_B}, except that it corresponds \ac{D-BLMMSE-DR}, and define $\setS'_{\textrm{D},k} = \{s_{l_{\textrm{D},k}^{(i)}}\}_{i=1}^{\nu}$. Then, the data detection problem in \eqref{eq:Doubly ML} simplifies to
\begin{align}\label{eq:Doubly ML (2)}
    \hat{\u}_{\textrm{D-ML}} = \argmin_{\u \in \prod_{k=1}^{K} \setS'_{\textrm{D},k}} \, \sum_{m=1}^{2M} \mathrm{log} \big(\Phi(\bar{\mu}_{\tilde{\y}|\x,m})\big).
\end{align}
As a result of the reduced search space, the total computational complexity decreases from $\mathcal{O}(MN^2L^K)$ to $\mathcal{O}(MN^2\nu^K)$. In our numerical results in Section~\ref{sec:NR}, we consider \ac{D-ML} with reduced search space specifying $\nu \in \{1, \ldots, L\}$, where the extreme cases $\nu=1$ and $\nu=L$ correspond to \ac{D-BLMMSE-DR} and full-search \ac{D-ML}, respectively.

\section{Complexity Analysis and Numerical Results} \label{sec:CCA&NR}

In this section, we first analyze the computational complexity of the proposed methods and then present numerical results to evaluate their performance.

\subsection{Complexity Analysis}\label{sec:CCA}

The computational complexity of the proposed methods is provided in Table~\ref{Table:DD}. We distinguish between complexity per channel, accounting for the computations required for a given channel realization, and complexity per data symbol vector.

\smallskip

\textit{\textbf{Data detection with full-resolution \acp{ADC}.}} The complexity of \ac{BLMMSE-DR} in Section~\ref{sec:DD1_A} is $\mathcal{O}(M^3+MN^2)$ per channel due to matrix inversion and multiplications for obtaining the soft estimates, and $\mathcal{O}(KL)$ per data symbol vector due to the minimum-distance mapping in \eqref{eq:l_k^star}. \ac{ML} in Section~\ref{sec:DD1_B} requires computing the statistics of the received signal for all the $L^K$ candidate data symbol vectors for each channel realization, which yields a per channel complexity of $\mathcal{O}(MN^2L^K)$. It then performs an exhaustive search over the same $L^K$ candidates, with complexity per data symbol vector of $\mathcal{O}(M^3L^K)$. The full-search \ac{ML-DR} method in Section~\ref{sec:DD1_B} evaluates the likelihood function in \eqref{eq:zeta_MLDR} for all the $L^K$ candidates. For each candidate, obtaining the symbol-dependent statistics has complexity of $\mathcal{O}(MN^2)$, while evaluating \eqref{eq:zeta_MLDR} with the full covariance matrix $\mathbf{\Sigma}_{{\tilde{\y}}^{\textrm{(DR)}}|{\x}}$ additionally requires matrix inversion and log-determinant computation with complexity of $\mathcal{O}(M^3)$. Therefore, the total complexity of full-search \ac{ML-DR} is $\mathcal{O}\big((M^3+MN^2)L^K\big)$. As discussed in Section~\ref{sec:DD1_B}, neglecting the off-diagonal elements of the covariance matrix reduces the total complexity to $\mathcal{O}(MN^2L^K)$, whereas restricting the search space to $\nu^K$ candidates identified by \ac{BLMMSE-DR}, with $\nu\leq L$, further reduces the complexity to $\mathcal{O}(MN^2\nu^K)$.

\smallskip

\textit{\textbf{Data detection with 1-bit \acp{ADC}.}} The complexity of \ac{D-BLMMSE-DR} in Section~\ref{sec:DD2_A} is the same as that of \ac{BLMMSE-DR}. The full-search \ac{D-ML} method in Section~\ref{sec:DD2_B} has total complexity of $\mathcal{O}(MN^2L^K)$, as it neglects the off-diagonal elements of $\mathbf{\Sigma}_{\tilde{\n}}(\x)$. Similar to \ac{ML-DR}, the complexity reduces to $\mathcal{O}(MN^2\nu^K)$ when the search space is restricted to $\nu^K$ candidates identified by \ac{D-BLMMSE-DR}.

\smallskip

Owing to their moderate complexity, the proposed soft-estimation-based method can support large $K$. On the other hand, the \ac{ML}-based methods are mainly suitable for small-to-moderate $K$ and large $M$.

\subsection{Numerical Results}\label{sec:NR}

In this section, we evaluate the data detection methods proposed in Sections~\ref{sec:DD1}~and~\ref{sec:DD2} in terms of \ac{SER} for the considered point-to-point massive \ac{MIMO} system employing 1-bit \acp{DAC} at the transmitter. To the best of our knowledge, the literature does not provide direct data detection methods tailored to this setting. Therefore, we introduce a two-stage data detection method, referred to as \textit{\ac{two-stage hoML}}, as a baseline. This method, described for convenience in Appendix~\ref{sec:App C}, first estimates the output of the 1-bit \acp{DAC} by means of the homotopy algorithm in \cite{Sha21} and then applies full-search \ac{ML}-based data detection to the resulting estimate.

Without loss of generality, we assume far-field propagation and generate the channel matrix $\H$ based on the discrete physical channel model described in~\cite{Sey02}. We assume that both the transmitter and receiver are equipped with a \ac{ULA} with half-wavelength antenna spacing. The arrays are positioned such that their broadside directions are aligned. A cluster of $10^2$ scatterers is placed between the transmitter and receiver, giving rise to as many propagation paths. At both ends, the scatterers are confined within an angular spread of $\frac{\pi}{6}$. The channels are normalized such that their elements have unit variance. The following \ac{SER} results are obtained by averaging over $10^2$ independent channel and \ac{AWGN} realizations, as well as $10^4$ independent data symbol vectors drawn from the 16-\ac{QAM} constellation. Furthermore, for each realization of $\H$, the precoding matrix $\W$ is set to comprise the $K$ principal right singular vectors of $\H$. For clarity, the values of the relevant system parameters, i.e., number of transmit and receive antennas $N$ and $M$, number of streams $K$, SNR $\rho$, and dither power $\sigma^{2}$, are indicated at the top of each figure. We note that the number of receive antennas is chosen to be small (i.e., $M=16$) for full-resolution \acp{ADC} and large (i.e., $M=128$) for 1-bit \acp{ADC}.

\begin{figure}[t!]
\centering
\begin{tikzpicture}[>=latex]

\begin{axis}[
	width=8cm,
	height=6.5cm,
	xmin=-40, xmax=0,
	ymin=1e-4, ymax=1,
    xlabel={$\sigma^2$ [dBm]},
    ylabel={SER},
    xlabel near ticks,
    ylabel near ticks,
    xtick={-40,-30,-20,-10,0},
    xticklabels={-10,0,10,20,30},
	ymode=log,
legend style={
    at={(1.16,0.02)},
    anchor=south east,
    font=\scriptsize,
    inner sep=1pt,
    fill opacity=0.75,
    draw opacity=1,
    text opacity=1,
},
    legend cell align=left,
	grid=both,
	x label style={font=\footnotesize},
	y label style={font=\footnotesize},
	ticklabel style={font=\footnotesize},
    clip marker paths=true,
    title={$N = 128, M = 16$, $K = 2$, $\rho=5$~dB},
 title style={font=\scriptsize, yshift=-2mm},
]

\addplot[very thick, blue, dashed]
table [x=Delta, y=SER_BLMMSE, col sep=comma] 
{Figs/Data/SER_ML_no_dither.txt};
\addlegendentry{BLMMSE}

\addplot[very thick, blue]
table [x=Delta, y=SER_BLMMSE_no_dith, col sep=comma] 
{Figs/Data/SER_ML_no_dither.txt};
\addlegendentry{BLMMSE-DR}

\addplot[very thick, black, dashed]
table [x=Delta, y=SER_ML, col sep=comma] 
{Figs/Data/SER_ML_no_dither.txt};
\addlegendentry{ML}

\addplot[ very thick, dashdotted, black, mark=o, mark options={fill=white,solid},mark layer=foreground,mark size=1.5pt,mark repeat=10] 
table [x=Delta, y=SER_ML_no_dith, col sep=comma] 
{Figs/Data/SER_ML_no_dither.txt};
\addlegendentry{\ac{ML-DR} ($\nu=L$), full $\mathbf{\Sigma}_{\tilde{\y}|\x}$}

\addplot[very thick, mark=square*,mark options={fill=white},mark layer=foreground, black, mark size=1.5pt,mark repeat=10]
table [x=Delta, y=SER_ML_no_dith_diag_Cov, col sep=comma] 
{Figs/Data/SER_ML_no_dither.txt};
\addlegendentry{\ac{ML-DR} ($\nu=L$)}

\addplot[very thick, dotted, mark=triangle*,mark options={fill=white,solid},mark layer=foreground, red, mark size=2.5pt,mark repeat=10]
table [x=Delta, y=SER_multi_ML, col sep=comma] 
{Figs/Data/SER_ML_no_dither.txt};
\addlegendentry{\ac{two-stage hoML}}

\end{axis}

\end{tikzpicture}
\caption{Full-resolution \acp{ADC}: \ac{SER} versus dither power.}\label{Fig:main_SER}
\end{figure}
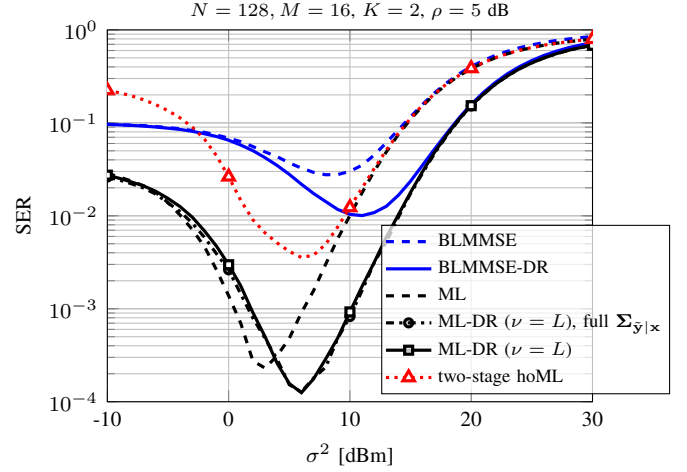

\subsubsection{Data detection with full-resolution \acp{ADC}}

Fig.~\ref{Fig:main_SER} plots the \ac{SER} versus the dither power $\sigma^{2}$ for fixed \ac{SNR} $\rho=5$~dB, with $K=2$ streams. All the methods exhibit an optimal dither power: for low $\sigma^2$, data symbols with the same phase produce nearly identical 1-bit quantized precoded signals, whereas for high $\sigma^2$ the Gaussian dither dominates. The proposed \ac{ML}-based methods, namely \ac{ML} and \ac{ML-DR}, outperform the soft-estimation-based methods \ac{BLMMSE} and \ac{BLMMSE-DR} by more than two orders of magnitude, since they directly exploit the statistics of the received signal, albeit at the cost of significantly higher computational complexity. For the soft-estimation-based methods, dither removal is always beneficial: \ac{BLMMSE-DR} outperforms \ac{BLMMSE} because subtracting the known dither-induced bias before linear combining reduces the \ac{MSE} of the soft estimates. On the other hand, the benefits of dither removal in the \ac{ML}-based methods only manifest after approximately $\sigma^2=3$~dBm. As discussed in Section~\ref{sec:DD1_B}, the effective noise $\n$ contains the dither-induced term, the quantization distortion $\p_{\tx}$ from the \acp{DAC}, and the \ac{AWGN}. For low-to-moderate $\sigma^2$, the dither-induced term partly mitigates the effect of $\p_{\tx}$, which preserves the accuracy of the likelihood function in \eqref{eq:zeta(x)} used by \ac{ML}. In contrast, \ac{ML-DR} removes this term, making the effect of $\p_{\tx}$ more pronounced and reducing the accuracy of the likelihood function in \eqref{eq:zeta_MLDR}. For larger $\sigma^2$, however, $\p_{\tx}$ becomes weakly correlated and the known dither-induced term dominates, so removing it improves the effective-noise statistics and allows \ac{ML-DR} to outperform \ac{ML}. At their respective optimal $\sigma^2$, \ac{ML-DR} achieves about a $2\times$ \ac{SER} reduction over \ac{ML}. Moreover, using only the diagonal elements of $\mathbf{\Sigma}_{\tilde{\y}|\x}$ in \ac{ML-DR} yields nearly the same performance as using the full covariance matrix. Since this approximation causes negligible loss while significantly reducing the computational complexity, the remaining figures consider \ac{ML-DR} with diagonal $\mathbf{\Sigma}_{\tilde{\y}|\x}$. For comparison, we also include \ac{two-stage hoML} from Appendix~\ref{sec:App C}. Although it outperforms \ac{BLMMSE-DR}, it is significantly more complex and remains inferior to the proposed \ac{ML}-based methods because its two-stage structure can introduce error propagation.

\begin{figure}[t!]
\centering
\begin{tikzpicture}[>=latex]

\begin{axis}[
	width=8cm,
	height=6.5cm,
	xmin=-40, xmax=0,
	ymin=1e-4, ymax=1,
    xlabel={$\sigma^2$ [dBm]},
    ylabel={SER},
    xlabel near ticks,
    ylabel near ticks,
    xtick={-40,-30,-20,-10,0},
    xticklabels={-10,0,10,20,30},
	ymode=log,
    legend style={at={(0.98,0.02)}, anchor=south east},
	legend style={font=\scriptsize, inner sep=1pt, fill opacity=0.75, draw opacity=1, text opacity=1},
    legend cell align=left,
	grid=both,
	x label style={font=\footnotesize},
	y label style={font=\footnotesize},
	ticklabel style={font=\footnotesize},
    clip marker paths=true,
    title={$N = 128, M = 16$, $K = 2$, $\rho=5$~dB},
 title style={font=\scriptsize, yshift=-2mm},
]

\addplot[very thick, mark=square*,mark options={fill=white},mark layer=foreground, brown, mark size=1.5pt,mark repeat=10]
table [x=Delta, y=lcML_nu1, col sep=comma] 
{Figs/Data/SER_lcML_vs_nu.txt};
\addlegendentry{\ac{ML-DR} ($\nu=1$)}

\addplot[very thick, mark=square*,mark options={fill=white},mark layer=foreground, black, mark size=1.5pt,mark repeat=10]
table [x=Delta, y=lcML_nu3, col sep=comma] 
{Figs/Data/SER_lcML_vs_nu.txt};
\addlegendentry{\ac{ML-DR} ($\nu=3$)}

\addplot[very thick, mark=square*,mark options={fill=white},mark layer=foreground, teal, mark size=1.5pt,mark repeat=10]
table [x=Delta, y=lcML_nu5, col sep=comma] 
{Figs/Data/SER_lcML_vs_nu.txt};
\addlegendentry{\ac{ML-DR} ($\nu=5$)}

\addplot[very thick, mark=square*,mark options={fill=white},mark layer=foreground, blue, mark size=1.5pt,mark repeat=10]
table [x=Delta, y=lcML_nuL, col sep=comma] 
{Figs/Data/SER_lcML_vs_nu.txt};
\addlegendentry{\ac{ML-DR} ($\nu=L$)}








\end{axis}

\end{tikzpicture}
\caption{Full-resolution \acp{ADC}: SER versus dither power.}\label{Fig:SER_vs_nu}
\end{figure}
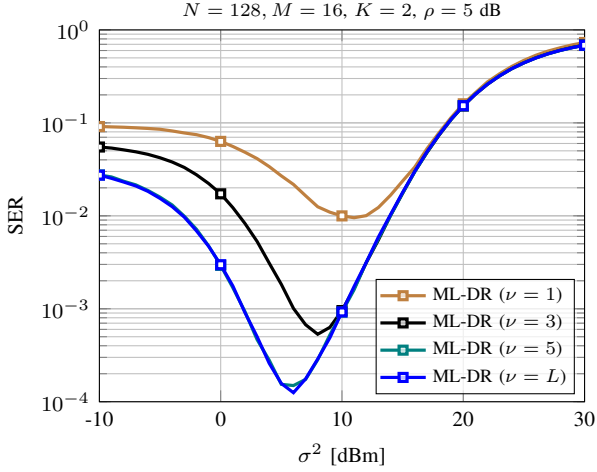

Considering the same setting of Fig.~\ref{Fig:main_SER}, Fig.~\ref{Fig:SER_vs_nu} compares the performance of \ac{ML-DR} for different values of $\nu$. As $\nu$ increases, the minimum \ac{SER} decreases and shifts to the left. Remarkably, the performance with reduced search space and $\nu = 5$ is indistinguishable from that with full search (i.e., $\nu = L$). Furthermore, comparing Figs.~\ref{Fig:main_SER} and~\ref{Fig:SER_vs_nu}, we observe that \ac{ML-DR} with $\nu=3$ already outperforms \ac{two-stage hoML}. Fig.~\ref{Fig:SER_vs_K} plots the minimum \ac{SER}, obtained via a linear search over the dither power $\sigma^2$, versus the number of streams $K$. We observe that all the data detection methods converge to nearly the same \ac{SER} as $K$ grows. In fact, for fixed $N$ and $M$, increasing $K$ boosts the inter-stream interference, which in turn reduces the beneficial impact of dithering at the transmitter, as noted in \cite{Sax20}. For the remaining results with full-resolution \acp{ADC}, we focus on \ac{ML-DR}.

\begin{figure}[t!]
\centering
\begin{tikzpicture}[>=latex]

\begin{axis}[
	width=8cm,
	height=6.5cm,
	xmin=3, xmax=8,
	ymin=1e-3, ymax=0.2,
    xlabel={$K$},
    ylabel={Minimum SER over $\sigma^2$},
    xlabel near ticks,
    ylabel near ticks,
    xtick={3,4,5,6,7,8},
    xticklabels={3,4,5,6,7,8},
	ymode=log,
    legend style={at={(0.98,0.02)}, anchor=south east},
	legend style={font=\scriptsize, inner sep=1pt, fill opacity=0.75, draw opacity=1, text opacity=1},
    legend cell align=left,
	grid=both,
	x label style={font=\footnotesize},
	y label style={font=\footnotesize},
	ticklabel style={font=\footnotesize},
    clip marker paths=true,
    title={$N = 128, M = 16$, $\rho=5$~dB},
 title style={font=\scriptsize, yshift=-2mm},
]


\addplot[very thick, blue, dashed]
table [x=KK, y=SER_BLMMSE_K, col sep=comma] 
{Figs/Data/minSER_vs_K.txt};
\addlegendentry{BLMMSE}

\addplot[very thick, blue]
table [x=KK, y=SER_BLMMSE_K_no_dith, col sep=comma] 
{Figs/Data/minSER_vs_K.txt};
\addlegendentry{BLMMSE-DR}



\addplot[very thick, mark=square*,mark options={fill=white},mark layer=foreground, black, mark size=1.5pt]
table [x=KK, y=SER_Heu_low_ML_Nheu3_K, col sep=comma]
{Figs/Data/minSER_vs_K.txt};
\addlegendentry{ML-DR ($\nu=3$)}


\end{axis}

\end{tikzpicture}
\caption{Full-resolution \acp{ADC}: minimum \ac{SER} over $\sigma^{2}$ versus number of data streams.}\label{Fig:SER_vs_K}
\end{figure}
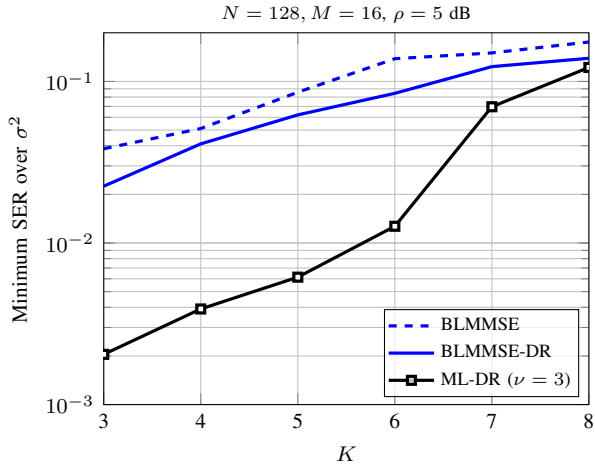

\begin{figure}[t!]
\centering
\begin{tikzpicture}[>=latex]

\begin{axis}[
	width=8cm,
	height=6.5cm,
	xmin=4, xmax=64,
	ymin=1e-6, ymax=0.2,
    xlabel={$M$},
    ylabel={Minimum SER over $\sigma^2$},
    xlabel near ticks,
    ylabel near ticks,
    xtick={4,8,16,32,64},
    xticklabels={4,8,16,32,64},
	ymode=log,
    legend style={at={(0.98,0.98)}, anchor=north east},
	legend style={font=\scriptsize, inner sep=1pt, fill opacity=0.75, draw opacity=1, text opacity=1},
    legend cell align=left,
	grid=both,
	x label style={font=\footnotesize},
	y label style={font=\footnotesize},
	ticklabel style={font=\footnotesize},
    clip marker paths=true,
    title={$N \in \{128, 256\}$, $K = 3$, $\rho = 5$~dB},
 title style={font=\scriptsize, yshift=-2mm},
]

\addplot[very thick, mark=square*,mark options={fill=white},mark layer=foreground, black, mark size=1.5pt]
table [x=M, y=minSER_N256, col sep=comma] 
{Figs/Data/minSER_vs_M.txt};
\addlegendentry{\ac{ML-DR}}

\addplot[
    very thick,
    dotted,
    red,
    mark=triangle*,
    mark options={fill=white,solid}, mark size=2.5pt
]
table [x=M, y=minSER_multi_N256, col sep=comma] 
{Figs/Data/minSER_vs_M.txt};
\addlegendentry{\ac{two-stage hoML}}

\addplot[very thick, mark=square*,mark options={fill=white},mark layer=foreground, black, mark size=1.5pt]
table [x=M, y=minSER, col sep=comma] 
{Figs/Data/minSER_vs_M.txt};

\addplot[very thick, mark=square*,mark options={fill=white},mark layer=foreground, black, mark size=1.5pt]
table [x=M, y=minSER_nu4, col sep=comma] 
{Figs/Data/minSER_vs_M.txt};

\addplot[very thick, mark=square*,mark options={fill=white},mark layer=foreground, black, mark size=1.5pt]
table [x=M, y=minSER_nu4_N256, col sep=comma] 
{Figs/Data/minSER_vs_M.txt};

\addplot[
    very thick,
    dotted,
    red,
    mark=triangle*,
    mark options={fill=white,solid}, mark size=2.5pt
]
table [x=M, y=minSER_multi, col sep=comma] 
{Figs/Data/minSER_vs_M.txt};

  \draw[
    black, thin, 
  ] (axis cs:16,2.1e-3) ellipse [x radius=10pt, y radius=23pt];

\node[anchor=center,font=\scriptsize] at (axis cs:16,3e-2) {$N = 128$};

\draw[
  black, thin, 
  ->,
] (axis cs:54,6e-4) -- (axis cs:54,1.3e-3)
  node[pos=0, anchor=north, font=\scriptsize] {$\nu = 3$};

\draw[
  black, thin, 
  ->,
] (axis cs:42,5e-4) -- (axis cs:42,2.1e-4)
  node[pos=0, anchor=south, font=\scriptsize] {$\nu = 4$};

 \draw[
  black
] (axis cs:16,2.3e-5) ellipse [x radius=10pt, y radius=22pt];

\node[anchor=center,font=\scriptsize] at (axis cs:16,2e-6) {$N = 256$};

\draw[
  black, thin, 
  ->,
  line width=0.2pt,
  >=Latex
] (axis cs:54,9e-6) -- (axis cs:54,2.5e-5)
  node[pos=0, anchor=north, font=\scriptsize] {$\nu = 3$};

\draw[
  black, thin,
  ->,
] (axis cs:42,8e-6) -- (axis cs:42,3.1e-6)
  node[pos=0, anchor=south, font=\scriptsize] {$\nu = 4$};

\end{axis}

\end{tikzpicture}
\caption{Full-resolution \acp{ADC}: minimum \ac{SER} over $\sigma^{2}$ versus number of receive antennas.}\label{Fig:SER_vs_M}
\end{figure}
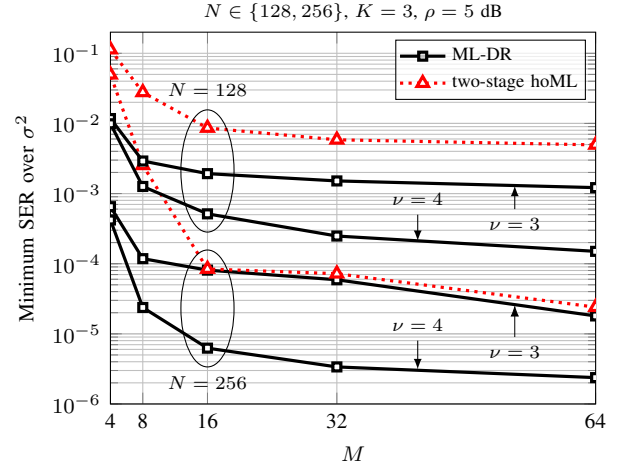

Fig.~\ref{Fig:SER_vs_M} illustrates the minimum \ac{SER} over $\sigma^2$ versus the number of receive antennas $M$, considering $K =3$ streams. For moderate-to-high values of $M$, \ac{ML-DR} with $\nu=4$ achieves up to approximately $30 \times$ lower \ac{SER} than \ac{two-stage hoML} for $N=128$, and up to approximately $20 \times$ lower \ac{SER} for $N=256$. Both \ac{ML-DR} and \ac{two-stage hoML} exhibit decreasing \ac{SER} with $M$ due to the improved spatial resolution. However, for large $M$, the \ac{SER} tends to saturate because the quantization distortion introduced by the 1-bit \acp{DAC} at the transmitter cannot be compensated by further increasing $M$. By contrast, increasing the number of transmit antennas from $N=128$ to $N=256$ yields \ac{SER} improvements of up to two orders of magnitude. For \ac{ML-DR}, this is because a larger $N$ improves the Gaussian approximation of the effective noise $\n$ in Section~\ref{sec:DD1_B}, thereby enhancing the \ac{ML} data detection in \eqref{eq:ML-DR}. For \ac{two-stage hoML}, the higher spatial resolution at the transmitter substantially improves the \ac{ML} data detection in the second stage in \eqref{eq:hoML} (see Appendix~\ref{sec:App C}).

\begin{figure}[t!]
\centering
\begin{tikzpicture}[>=latex]

\begin{axis}[
	width=8cm,
	height=6.5cm,
	xmin=-40, xmax=0,
	ymin=1e-3, ymax=1,
    xlabel={$\sigma^2$ [dBm]},
    ylabel={SER},
    xlabel near ticks,
    ylabel near ticks,
    xtick={-40,-30,-20,-10,0},
    xticklabels={-10,0,10,20,30},
	ymode=log,
    legend style={at={(0.98,0.02)}, anchor=south east},
	legend style={font=\scriptsize, inner sep=1pt, fill opacity=0.75, draw opacity=1, text opacity=1},
    legend cell align=left,
	grid=both,
	x label style={font=\footnotesize},
	y label style={font=\footnotesize},
	ticklabel style={font=\footnotesize},
    clip marker paths=true,
    title={$N = 128, M = 16$, $K = 3$, $\rho = 5$~dB},
 title style={font=\scriptsize, yshift=-2mm},
]

\addplot[very thick, mark=square*,mark options={fill=white},mark layer=foreground, black, mark size=1.5pt,mark repeat=10]
table [x=Delta, y=SER_lcML, col sep=comma] 
{Figs/Data/SER_imperfect_CSI.txt};
\addlegendentry{ML-DR ($\nu = 3$)}

\addplot[very thick, mark=square*,mark options={solid,fill=white},mark layer=foreground, black, mark size=1.5pt,mark repeat=10,forget plot]
table [x=Delta, y=SER_lcML_ch_es, col sep=comma] 
{Figs/Data/SER_imperfect_CSI.txt};

\addplot[
    very thick,
    dotted,
    red,
    mark=triangle*,
    mark options={fill=white,solid}, mark size=2.5pt,mark repeat=10]
table [x=Delta, y=SER_multi_ML, col sep=comma] 
{Figs/Data/SER_imperfect_CSI.txt};
\addlegendentry{two-stage hoML}

\addplot[
    very thick, dotted, red, mark=triangle*,mark options={fill=white,solid}, mark size=2.5pt,mark repeat=10,forget plot]
table [x=Delta, y=SER_multi_ML_imperfect_CSI, col sep=comma] 
{Figs/Data/SER_imperfect_CSI.txt};

\node[anchor=center,font=\scriptsize] at (axis cs:-20,6e-1) {imperfect CSI};
\node[anchor=center,font=\scriptsize] at (axis cs:-35,2e-3) {perfect CSI};

\draw[->, thin] (axis cs:-25,6e-1) -- (axis cs:-30,3.4e-1);
\draw[->, thin] (axis cs:-25,6e-1) -- (axis cs:-30,1.7e-2);
\draw[->, thin] (axis cs:-35,2.5e-3) -- (axis cs:-30,5.5e-2);
\draw[->, thin] (axis cs:-35,2.5e-3) -- (axis cs:-30,9.5e-3);

\end{axis}

\end{tikzpicture}
\caption{Full-resolution \acp{ADC}: \ac{SER} versus dither power under perfect and imperfect \ac{CSI} at the receiver (with pilot length $\tau=131$).}\label{Fig:SER_imp.CSI}
\end{figure}
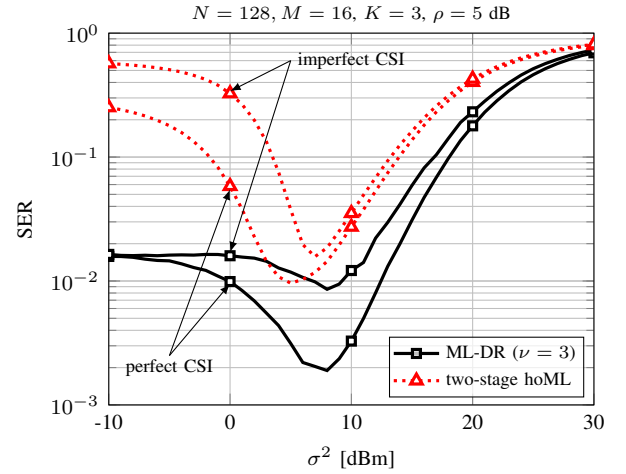

While the proposed methods are designed for perfect \ac{CSI}, Fig.~\ref{Fig:SER_imp.CSI} shows the \ac{SER} versus the dither power $\sigma^{2}$ for fixed \ac{SNR} $\rho=5$~dB under imperfect \ac{CSI} at the receiver, with $K=3$ streams. Here, imperfect \ac{CSI} only affects the data detection at the receiver, whereas the transmitter is assumed to have perfect \ac{CSI} for constructing the precoding matrix $\W$. To estimate the channel at the receiver, orthogonal antenna-specific Zadoff-Chu pilot sequences \cite{Hyd17} are transmitted from the transmit antennas, with pilot length $\tau = 131 \geq N$ to ensure orthogonality. The pilots are transmitted without precoding and are dithered prior to the 1-bit \acp{DAC}. The receiver (with full-resolution \acp{ADC}), then performs \ac{LMMSE} channel estimation by treating each transmit antenna as an independent pilot source. The performance of \ac{ML-DR} with $\nu = 3$ degrades under imperfect \ac{CSI}, but still remains superior to that of \ac{two-stage hoML} under perfect \ac{CSI}, although the latter inherently relies on full-search \ac{ML} data detection as in \eqref{eq:hoML}.

\begin{figure}[t!]
\centering
\begin{tikzpicture}[>=latex]
\begin{axis}[
	width=8cm, height=6.5cm,
	xmin=-10, xmax=20,
	ymin=-10, ymax=30,
	zmin=1e-3, zmax=1,
    zmode=log,
	view={75}{30},
	xlabel={$\rho$ [dB]},
	ylabel={$\sigma^{2}$ [dBm]},
	zlabel={SER},
	x label style={at={(axis cs:5,-20,1e-3)}, anchor=north},
	y label style={at={(axis cs:30,10,1e-3)}, anchor=north},
	label style={font=\footnotesize},
    xtick={-10,0,...,20},
    ytick={-10,0,...,30},
    ztick={1e-3,1e-2,1e-1,1},
	yticklabels={$\hspace{2mm} -10$, $0$, $10$, $20$, $30$},
	zticklabels={$10^{-3}$,$10^{-2}$,$10^{-1}$,$10^{0}$},
	ticklabel style={font=\footnotesize, yshift=-2mm},
	%
    %
	title={$N = 128$, $M = 16$, $K = 3$},
	title style={font=\scriptsize, yshift=-2mm},
	grid=major,
	colormap/jet,
	mesh/rows=16, 
	mesh/cols=41, 
]

\addplot3[surf, opacity=0.9]
table {anc/SER_3D.dat};

\end{axis}

\end{tikzpicture}
\caption{Full-resolution \acp{ADC}: \ac{SER} obtained with \ac{ML-DR} ($\nu = 3$) versus \ac{SNR} and dither power.}\label{Fig:SER_3D}
\end{figure}
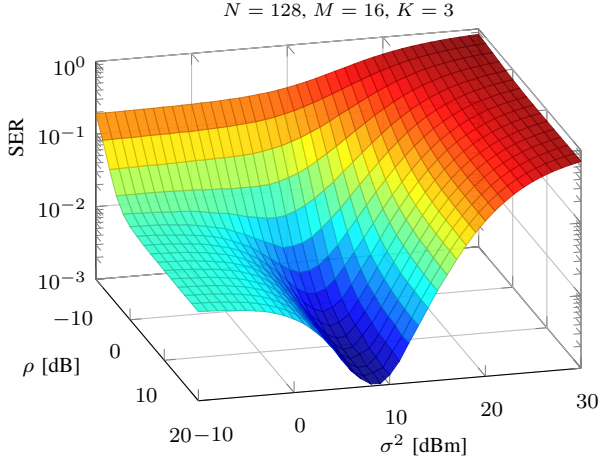

Returning to perfect \ac{CSI}, Fig.~\ref{Fig:SER_3D} presents a 3D plot of the \ac{SER} achieved by \ac{ML-DR} with $\nu=3$ versus the \ac{SNR} $\rho$ and dither power $\sigma^2$, with $K=3$~streams. At low \ac{SNR}, the \ac{AWGN} is dominant, and thus dithering is not beneficial as it additionally perturbs the transmit signal. After approximately $\rho = 0$~dB, we clearly observe an optimal operating point as a function of the dither power. However, after approximately $\rho = 8$~dB, increasing $\rho$ no longer improves the \ac{SER} at the optimal $\sigma^2$, since the quantization distortion at the transmitter remains the dominant limiting factor. This behavior is also illustrated in Fig.~\ref{Fig:SER_vs_rho}, which plots the \ac{SER} versus the \ac{SNR} $\rho$ for fixed dither power $\sigma^2 = 8$~dBm. Expanding the search space by increasing $\nu$ from $\nu=3$ to $\nu=4$ provides a $3 \times$ \ac{SER} reduction, corresponding to a gain of nearly two orders of magnitude over \ac{two-stage hoML} at moderate-to-high~\ac{SNR}.

\begin{figure}[t!]
\centering
\begin{tikzpicture}[>=latex]

\begin{axis}[
	width=8cm,
	height=6.5cm,
	xmin=0, xmax=30,
	ymin=1e-4, ymax=6e-2,
    xlabel={$\rho$ [dB]},
    ylabel={SER},
    xlabel near ticks,
    ylabel near ticks,
    xtick={0,10,20,30},
    xticklabels={0,10,20,30},
	ymode=log,
    legend style={at={(0.98,0.98)}, anchor=north east},
	legend style={font=\scriptsize, inner sep=1pt, fill opacity=0.75, draw opacity=1, text opacity=1},
    legend cell align=left,
	grid=both,
	x label style={font=\footnotesize},
	y label style={font=\footnotesize},
	ticklabel style={font=\footnotesize},
    clip marker paths=true,
    title={$N = 128$, $M = 16$, $K = 3$, $\sigma^{2} = 8$~dBm},
 title style={font=\scriptsize, yshift=-2mm},
]

\addplot[very thick, mark=square*,mark options={fill=white},mark layer=foreground, black, mark size=1.5pt,mark repeat=5]
table [x=rho_dB, y=ELCML_Nhe3, col sep=comma] 
{Figs/Data/SER_vs_rho_at_delta_smooth.txt};
\addlegendentry{\ac{ML-DR}}

\addplot[very thick, mark=square*,mark options={fill=white},mark layer=foreground, black, mark size=1.5pt,mark repeat=5,forget plot]
table [x=rho_dB, y=ELCML_Nhe4, col sep=comma, forget plot] 
{Figs/Data/SER_vs_rho_at_delta_smooth.txt};

\addplot[
    very thick,
    dotted,
    red,
    mark=triangle*,
    mark options={fill=white,solid}, mark size=2.5pt,mark repeat=5]
table [x=rho_dB, y=Multi_ML, col sep=comma] 
{Figs/Data/SER_vs_rho_at_delta_smooth.txt};
\addlegendentry{\ac{two-stage hoML}}

  \draw[
    black, thin,
    ->,
  ] (axis cs:25,2e-3) -- (axis cs:20,1.1e-3)
    node[pos=0, anchor=south,font=\scriptsize] {$\nu=3$};

  \draw[
    black, thin,
    ->,
  ] (axis cs:25,5e-4) -- (axis cs:20,3.2e-4)
    node[pos=0, anchor=south,font=\scriptsize] {$\nu=4$};

\end{axis}

\end{tikzpicture}
\caption{Full-resolution \acp{ADC}: \ac{SER} versus \ac{SNR}.}\label{Fig:SER_vs_rho}
\end{figure}
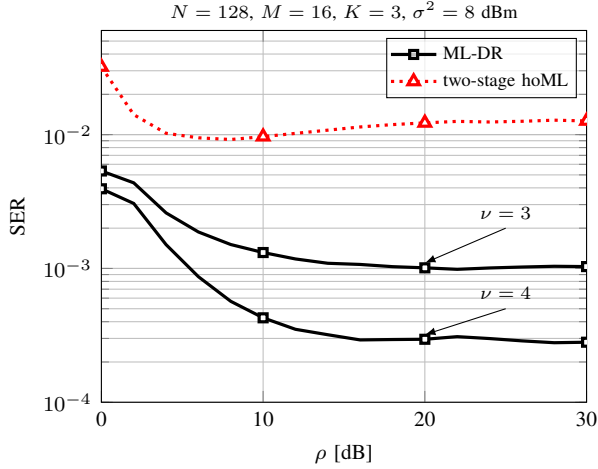

\begin{figure}[t!]
\centering
\begin{tikzpicture}[>=latex]

\begin{axis}[
	width=8cm,
	height=6.5cm,
	xmin=-40, xmax=0,
	ymin=1e-2, ymax=1,
    xlabel={$\sigma^2$ [dBm]},
    ylabel={SER},
    xlabel near ticks,
    ylabel near ticks,
    xtick={-40,-30,-20,-10,0},
    xticklabels={-10,0,10,20,30},
	ymode=log,
    legend style={at={(0.98,0.02)}, anchor=south east},
	legend style={font=\scriptsize, inner sep=1pt, fill opacity=0.75, draw opacity=1, text opacity=1},
    legend cell align=left,
	grid=both,
	x label style={font=\footnotesize},
	y label style={font=\footnotesize},
	ticklabel style={font=\footnotesize},
    clip marker paths=true,
    title={$M = N = 128$, $K = 2$, $\rho = 8$~dB},
 title style={font=\scriptsize, yshift=-2mm},
]
\addplot[very thick, blue, dashed]
table [x=Delta, y=SER_BLMMSE, col sep=comma] 
{Figs/Data/SER_vs_dither_ML_doubly_MN128_K2.txt};
\addlegendentry{D-BLMMSE}

\addplot[very thick, blue]
table [x=Delta, y=SER_BLMMSE_no_dith, col sep=comma] 
{Figs/Data/SER_vs_dither_ML_doubly_MN128_K2.txt};
\addlegendentry{D-BLMMSE-DR}

\addplot[very thick, mark=square*,mark options={fill=white},mark layer=foreground, black, mark size=1.5pt,mark repeat=5]
table [x=Delta, y=SER_ML, col sep=comma] 
{Figs/Data/SER_vs_dither_ML_doubly_MN128_K2.txt};
\addlegendentry{\ac{D-ML}}


\addplot[very thick, mark=square*,mark options={fill=white},mark layer=foreground, black, mark size=1.5pt,mark repeat=5,forget plot]
table [x=Delta, y=SER_Heu_low_comp, col sep=comma] 
{Figs/Data/SER_vs_dither_ML_doubly_MN128_K2.txt};

\addplot[
    very thick,
    dotted,
    red,
    mark=triangle*,
    mark options={fill=white,solid}, mark size=2.5pt,mark repeat=5]
table [x=Delta, y=SER_multi_ML, col sep=comma] 
{Figs/Data/SER_vs_dither_ML_doubly_MN128_K2.txt};
\addlegendentry{\ac{two-stage hoML}}


  \draw[
    black, thin, 
    ->,
  ] (axis cs:-35,7e-2) -- (axis cs:-30,4.5e-2)
    node[pos=0, anchor=south,font=\scriptsize] {$\nu = 3$};

    \draw[
    black, thin,
    ->,
  ] (axis cs:-35,3e-2) -- (axis cs:-30,1.4e-2)
    node[pos=0, anchor=south,font=\scriptsize] {$\nu = L$};

\end{axis}

\end{tikzpicture}
\caption{1-bit \acp{ADC}: \ac{SER} versus dither power.}\label{Fig:SER_vs_dither_doubly}
\end{figure}
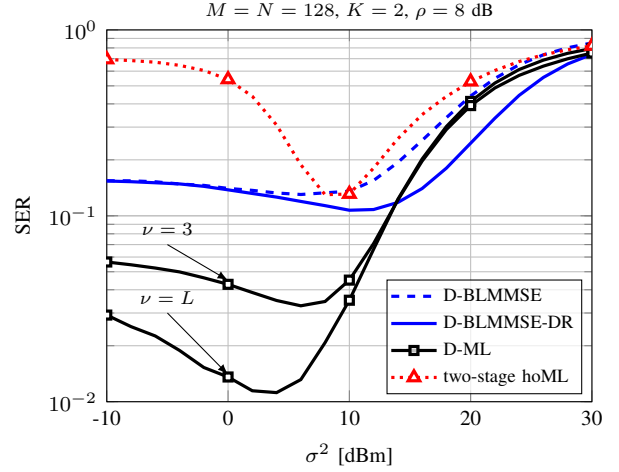

\subsubsection{Data detection with 1-bit \acp{ADC}}

Fig.~\ref{Fig:SER_vs_dither_doubly} plots the \ac{SER} versus the dither power $\sigma^2$. Full-search \ac{D-ML} achieves a $10\times$ \ac{SER} reduction over the other data detection methods at their respective optimal $\sigma^2$. This is because \ac{D-ML} performs direct \ac{ML} data detection based on the 1-bit quantized received signal $\r$ while properly exploiting the statistics of the received signal $\y$ (prior to the 1-bit \acp{ADC}), as described in Section~\ref{sec:DD2_B}. As a result, \ac{D-ML} avoids the error propagation inherent to the \ac{two-stage hoML} method. Furthermore, \ac{D-BLMMSE-DR} outperforms \ac{two-stage hoML}, despite the latter having much higher computational complexity.

Lastly, Fig.~\ref{Fig:SER_vs_rho_doubly} plots the \ac{SER} versus the \ac{SNR} $\rho$ for fixed dither power $\sigma^{2} = 2$~dBm. \ac{D-ML} with $\nu = 3$ already outperforms \ac{two-stage hoML} over the entire \ac{SNR} range. Moreover, the \ac{SER} achieved by \ac{D-ML} and \ac{D-BLMMSE-DR} tends to saturate as $\rho$ increases, for reasons similar to those discussed in Fig.~\ref{Fig:SER_3D}, and \ac{D-BLMMSE-DR} is outperformed by \ac{two-stage hoML} only for high $\rho$.

\begin{figure}[t!]
\centering
\begin{tikzpicture}[>=latex]

\begin{axis}[
	width=8cm,
	height=6.5cm,
	xmin=0, xmax=30,
	ymin=7e-3, ymax=1,
    xlabel={$\rho$ [dB]},
    ylabel={SER},
    xlabel near ticks,
    ylabel near ticks,
    xtick={0,10,20,30},
    xticklabels={0,10,20,30},
	ymode=log,
    legend style={at={(0.98,0.98)}, anchor=north east},
	legend style={font=\scriptsize, inner sep=1pt, fill opacity=0.75, draw opacity=1, text opacity=1},
    legend cell align=left,
	grid=both,
	x label style={font=\footnotesize},
	y label style={font=\footnotesize},
	ticklabel style={font=\footnotesize},
    clip marker paths=true,
    title={$M = N = 128$, $K = 2$, $\sigma^2 = 2$~dBm},
 title style={font=\scriptsize, yshift=-2mm},
]


\addplot[very thick, blue]
table [x=rhodB, y=SER_BLMMSE_dith_cancell, col sep=comma] 
{Figs/Data/SER_vs_rho_ML_doubly_MN128_K2.txt};
\addlegendentry{D-BLMMSE-DR}

\addplot[very thick, mark=square*,mark options={fill=white},mark layer=foreground, black, mark size=1.5pt,mark repeat=5,forget plot]
table [x=rhodB, y=SER_ML, col sep=comma] 
{Figs/Data/SER_vs_rho_ML_doubly_MN128_K2.txt};

\addplot[very thick, mark=square*,mark options={fill=white},mark layer=foreground, black, mark size=1.5pt,mark repeat=5]
table [x=rhodB, y=SER_Heu_low_comp, col sep=comma] 
{Figs/Data/SER_vs_rho_ML_doubly_MN128_K2.txt};
\addlegendentry{\ac{D-ML}}

\addplot[
    very thick,
    dotted,
    red,
    mark=triangle*,
    mark options={fill=white,solid}, mark size=2.5pt,mark repeat=5]
table [x=rhodB, y=SER_multi_ML, col sep=comma] 
{Figs/Data/SER_vs_rho_ML_doubly_MN128_K2.txt};
\addlegendentry{\ac{two-stage hoML}}


  \draw[
    black, thin, 
    ->,
  ] (axis cs:5,6e-2) -- (axis cs:10,4e-2)
    node[pos=0, anchor=south,font=\scriptsize] {$\nu = 3$};

  \draw[
    black, thin, 
    ->,
  ] (axis cs:5,2e-2) -- (axis cs:10,1.1e-2)
    node[pos=0, anchor=south,font=\scriptsize] {$\nu = L$};

\end{axis}

\end{tikzpicture}
\caption{1-bit \acp{ADC}: \ac{SER} versus \ac{SNR}.}\label{Fig:SER_vs_rho_doubly}
\end{figure}
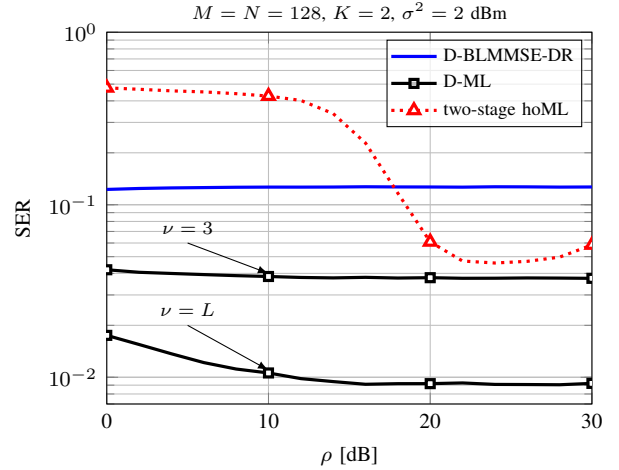

\begin{figure*}[t!]
\setcounter{equation}{59}
\begin{align}
& \C_{\tilde{\n}}^{(\mathrm{A},\mathrm{B})} \! = \! \rho\mathbb{E}\big[\mathrm{A}[\H\p_{\tx}]\mathrm{B}[\H\p_{\tx}]^\tran \! + \! \mathrm{A}[\H_{\G}\d]\mathrm{B}[\H\p_{\tx}]^\tran \! + \! \mathrm{A}[\H\p_{\tx}]\mathrm{B}[\H_{\G}\d]^\tran\big] \! + \! \frac{\rho\sigma^2}{2}\Lambda_{\mathrm{A},\mathrm{B}}\big(\H_{\G}\H_{\G}^\herm\big) \! + \! \frac{\delta_{\mathrm{A},\mathrm{B}}}{2}\I_M \in \Real^{M \times M}, \label{eq:Cnt_Re} \\
& \mathbb{E}_{\p_{\tx}}\big[\mathrm{A}[\p_{\tx}]\mathrm{B}[\p_{\tx}]^\tran\big] \! = \! \C_{\x_{\textrm{q}}}^{(\mathrm{A},\mathrm{B})} \! - \! {\P_1}^{(\mathrm{A},\mathrm{B})} \! - \! {\P_2}^{(\mathrm{A},\mathrm{B})} \! + \! \frac{\sigma^2}{2}\Lambda_{\mathrm{A},\mathrm{B}}\big(\G_{\tx}(\x)\G_{\tx}(\x)^\herm\big) \! + \! \mathrm{A}[\G_{\tx}(\x)\x]\mathrm{B}[\G_{\tx}(\x)\x]^\tran \in \mathbb{R}^{N\times N}, \label{eq:Cqx_re} \\
& [\C_{\x_{\textrm{q}}}^{( \mathrm{A}, \mathrm{B})}]_{n,m} \triangleq\mathbb{E}_{\x_{\textrm{q}}}\big[\mathrm{A}[\x_{\textrm{q}}]\mathrm{B}[\x_{\textrm{q}}]^\tran \big]_{n,m} =
    \begin{cases}
    \frac{\eta}{2}\big(1 - \erf \big(\frac{\mathrm{A}[x_n]}{\sigma} \big)\erf \big(\frac{\mathrm{B}[x_m]}{\sigma} \big)\big)\delta_{\mathrm{A},\mathrm{B}} + \frac{\eta}{2}\erf \big(\frac{\mathrm{A}[x_n]}{\sigma} \big)\erf \big(\frac{\mathrm{B}[x_m]}{\sigma} \big)& \ \textrm{if}~m=n, \\
    \frac{\eta}{2} \erf \big(\frac{\mathrm{A}[x_n]}{\sigma} \big)\erf \big(\frac{\mathrm{B}[x_m]}{\sigma} \big) & \ \textrm{otherwise}.
    \end{cases} \label{eq:Exq_n Exq_m}
\end{align}
\hrulefill \vspace{-2mm}
\end{figure*}

\section{Conclusions} \label{sec:Contr_Full}

In this paper, we studied data detection for a point-to-point massive \ac{MIMO} system with 1-bit quantized dithered linear precoding at the transmitter and either full-resolution or 1-bit \acp{ADC} at the receiver. Assuming that the dither vector applied at the transmitter is known at the receiver, we developed soft-estimation-based data detection methods with symbol-independent dither removal, namely \textit{\ac{BLMMSE-DR}} and \textit{\ac{D-BLMMSE-DR}} for full-resolution and 1-bit \acp{ADC}, respectively. Then, we introduced a new symbol-dependent linearization of the 1-bit quantized transmitted signal, based on which we derived \ac{ML}-based data detection methods that directly recover the data symbol vector from the received signal. Specifically, we proposed \textit{\ac{ML-DR}} for full-resolution \acp{ADC}, which incorporates symbol-dependent dither removal, and \textit{\ac{D-ML}} for 1-bit \acp{ADC}. For both methods, we developed low-complexity variants to mitigate the exponential complexity growth with the number of data streams. Numerical results in terms of \ac{SER} highlighted the critical role of the dither power and showed that the proposed \ac{ML}-based methods, together with their low-complexity variants, achieve significant gains over both the soft-estimation-based methods and the \ac{two-stage hoML} baseline. For full-resolution \acp{ADC}, \ac{ML-DR} consistently outperformed \ac{two-stage hoML} and achieved up to two orders of magnitude lower \ac{SER} than the soft-estimation-based methods, while dither removal provided a $2\times$ \ac{SER} reduction compared with the case without dither removal. For 1-bit \acp{ADC}, \ac{D-ML} achieved a $10\times$ \ac{SER} reduction over the other methods. Future work may consider extension to imperfect \ac{CSI}, the optimization of the dither power, and precoding design tailored to the proposed data detection methods.

\appendices 

\section{Derivation of $\C_{\x_{\textnormal{d}}\x_{\textnormal{q}}|\x}^{(\mathrm{A},\mathrm{B})}$} \label{sec:App A}

In this appendix, we derive the closed-form expression for the elements of
$\C_{\x_{\textrm{d}}\x_{\textrm{q}}|\x}^{(\mathrm{A},\mathrm{B})}$
defined in Section~\ref{sec:LQP&DD_B}. We begin by writing the $(n,m)$th element of
$\C_{\x_{\textrm{d}}\x_{\textrm{q}}|\x}^{(\mathrm{A},\mathrm{B})}$ as
\setcounter{equation}{46}
\begin{align}\label{eq:Cxdxq prime}
[\C_{\x_{\textrm{d}}\x_{\textrm{q}}|\x}^{(\mathrm{A}, \mathrm{B})}]_{n,m}
= \sqrt{\frac{\eta}{2}} \mathbb{E} \big[\mathrm{A}[x_{\textrm{d},n}]\sgn \big(\mathrm{B}[x_{\textrm{d},m}] \big)\big|\x\big].
\end{align}
For a given $\x$, define the real-valued random variables $\theta_{n}^{(\mathrm{A})} \triangleq \mathrm{A}[x_{\textrm{d},n}]$ and $\theta_{m}^{(\mathrm{B})} \triangleq \mathrm{B}[x_{\textrm{d},m}]$, with $\theta_{n}^{(\mathrm{A})}\sim\mathcal{N}(\mathrm{A}[x_n],\frac{\sigma^2}{2})$ and
$\theta_{m}^{(\mathrm{B})}\sim\mathcal{N}(\mathrm{B}[x_m],\frac{\sigma^2}{2})$. Hence,
\eqref{eq:Cxdxq prime} can be expressed as
\begin{align}
[\C_{\x_{\textrm{d}}\x_{\textrm{q}}|\x}^{(\mathrm{A},\mathrm{B})}]_{n,m}
&=
\sqrt{\frac{\eta}{2}}
\mathbb E\big[\theta_{n}^{(\mathrm{A})}\sgn(\theta_{m}^{(\mathrm{B})})|\x\big].
\label{eq:Cxdxq_compact}
\end{align}

For $n=m$ and $\mathrm{A}=\mathrm{B}$, we have $\theta_{n}^{(\mathrm{A})}=\theta_{m}^{(\mathrm{B})}$ and thus
$\theta_{n}^{(\mathrm{A})}\sgn(\theta_{m}^{(\mathrm{B})})=|\theta_{n}^{(\mathrm{A})}|$. Therefore, \eqref{eq:Cxdxq_compact} becomes
\begin{align}
[\C_{\x_{\textrm{d}}\x_{\textrm{q}}|\x}^{(\mathrm{A},\mathrm{B})}]_{n,n}
&=
\sqrt{\frac{\eta}{2}}
\mathbb E\big[|\theta_{n}^{(\mathrm{A})}|\,|\x\big]\\
&= \nonumber\sqrt{\frac{\eta}{2}} \bigg(
\sqrt{\frac{\sigma^2}{\pi}} 
\exp \bigg(-\bigg(\frac{\mathrm{A}[x_n]}{\sigma}\bigg)^2\bigg)\\
    & \phantom{=} \ 
+ \mathrm{A}[x_n] \erf \left(\frac{\mathrm{A}[x_n]}{\sigma}\right)
\bigg), \label{eq:Cxdxq_caseII}
\end{align}
where \eqref{eq:Cxdxq_caseII} follows from taking the expectation of the absolute value of a Gaussian random variable. Otherwise, if $n=m$ and $\mathrm{A}=\mathrm{B}$ do not simultaneously hold, the random variables $\theta_{n}^{(\mathrm{A})}$ and $\theta_{m}^{(\mathrm{B})}$ are independent. In this case, we have
\begin{align}
\mathbb E\big[\theta_{n}^{(\mathrm{A})}\sgn(\theta_{m}^{(\mathrm{B})})|\x\big] & =\mathbb E[\theta_{n}^{(\mathrm{A})}|\x]\mathbb E[\sgn(\theta_{m}^{(\mathrm{B})})|\x] \\
& = \mathrm{A}[x_n] \erf \left(\frac{\mathrm{B}[x_m]}{\sigma}\right),
\end{align}
which readily leads to
\begin{align}
[\C_{\x_{\textrm{d}}\x_{\textrm{q}}|\x}^{(\mathrm{A},\mathrm{B})}]_{n,m}
&=
\sqrt{\frac{\eta}{2}}
\mathrm{A}[x_n]
\erf \left(\frac{\mathrm{B}[x_m]}{\sigma}\right).
\label{eq:Cxdxq_case2}
\end{align}

\section{Derivations of the Mean and the Covariance of $\tilde{\n}$} \label{sec:App B}

In this appendix, we derive the mean and covariance of $\tilde{\n}$ defined in Section~\ref{sec:DD1_B}. Define $\H_{\G} \triangleq \H\G_{\tx}(\x) \in \Compl^{M\times N}$. Unless otherwise stated, all the expectations are taken over $\d$, $\p_{\tx}$, and $\z$, conditioned on $\x$.

Recalling the definition of $\mub_{\n}^{(\mathrm{A})}$ from Section~\ref{sec:DD1_B}, the mean of $\tilde{\n}$ is given by
\begin{align}
\boldsymbol{\mu}_{\tilde{\n}}& = \big[(\mub_{\n}^{(\Re)})^\tran, (\mub_{\n}^{(\Im)})^\tran\big],\label{eq:mu_n'} 
\end{align}
with
\begin{align}
\mub_{\n}^{(\Re)} &= \sqrt{\rho} \big(\Re[\H]\mathbb E\big[\Re[\p_{\tx}]\big] - \Im[\H]\mathbb E\big[\Im[\p_{\tx}]\big] \big), \label{eq:E[Re[zt]]}
     \\ \mub_{\n}^{(\Im)} &= \sqrt{\rho} \big(\Re[\H]\mathbb E\big[\Im[\p_{\tx}]\big] + \Im[\H]\mathbb E\big[\Re[\p_{\tx}]\big] \big),\label{eq:E[Im[zt]]}
\end{align}
and where, recalling the definition of $\p_{\tx}$ in \eqref{eq:xq (2)}, we have
\begin{align}
    \mathbb E\big[\mathrm{A}[\p_{\tx}]\big] &= \mathbb E\big[\mathrm{A}[\x_{\textrm{q}}]\big] -  \mathrm{A} \big[\G_{\tx}(\x)\x \big], \label{eq:Eqx_A}
\end{align} 
with
\begin{align}
   \mathbb E\big[\mathrm{A}[\x_{\textrm{q}}]\big] & =
    \sqrt{\frac{\eta}{2}} \erf\bigg(\frac{\mathrm{A}[\x]}{\sigma}\bigg). \label{eq:Exq}
\end{align} 

The second-order moment of $\tilde{\n}$ is given by 
\begin{align}\label{eq:Cn't}
    \C_{\tilde{\n}} \triangleq \mathbb{E} [\tilde{\n}{\tilde{\n}}^\tran] =\begin{bmatrix}
        \C_{\tilde{\n}}^{(\Re,\Re)} & \C_{\tilde{\n}}^{(\Re,\Im)}\\
        \C_{\tilde{\n}}^{(\Im,\Re)} & \C_{\tilde{\n}}^{(\Im,\Im)}
    \end{bmatrix} \in \mathbb{R}^{2M\times 2M},
    \end{align}
with specific term in \eqref{eq:Cnt_Re}--\eqref{eq:Exq_n Exq_m} at the top of the page. In \eqref{eq:Cnt_Re}, the cross-correlation matrix between $\d$ and $\p_{\tx}$ is given by
\setcounter{equation}{62}
\begin{align}
    \C_{\d\p_{\tx}}^{(\mathrm{A}, \mathrm{B})} &\triangleq\mathbb{E}_{\d,\p_{\tx}}\big[ \mathrm{A}[\d]\mathrm{B}[\p_{\tx}]^\tran\big] \\&= \C_{\x_{\textrm{d}}\x_{\textrm{q}}}^{( \mathrm{A}, \mathrm{B})} \! - \! \mathbb{E}\big[\mathrm{A}[\d]\mathrm{B} \big[\G_{\tx}(\x)\d \big]^\tran\big] \! - \! \mathrm{A}[\x]\mathbb{E}\big[\mathrm{B}[\x_{\textrm{q}}]\big]^\tran, \label{eq:Edqx_AB}
\end{align}
with $\mathbb{E}\big[\mathrm{B}[\x_{\textrm{q}}]\big]$ given in \eqref{eq:Exq} and where $\mathbb{E}\big[\mathrm{A}[\d]\mathrm{B} \big[\G_{\tx}(\x)\d \big]^\tran\big]$ can be derived based on $\mathbb{E}\big[\mathrm{A}[\d]\mathrm{B}[\d]^\tran\big] = \delta_{\mathrm{A},\mathrm{B}}\, \frac{\sigma^2}{2}\I_N$. Moreover, in \eqref{eq:Cqx_re}, we defined
\begin{align} \label{eq:P1}
    {\P_1}^{(\mathrm{A},\mathrm{B})} &\triangleq \mathbb{E}_{\d,\x_{\textrm{q}}}\big[ \mathrm{A}[\G_{\tx}(\x)\x_{\textrm{d}}]\mathrm{B}[\x_{\textrm{q}}]^\tran \big], \\\label{eq:P2}
    {\P_2}^{(\mathrm{A},\mathrm{B})} &\triangleq \mathbb{E}_{\d,\x_{\textrm{q}}}\big[\mathrm{A}[\x_{\textrm{q}}] \mathrm{B}[\G_{\tx}(\x)\x_{\textrm{d}}]^\tran \big],
\end{align}
where the $(n,m)$th element of $\mathbb{E}_{\d,\x_{\textrm{q}}}\big[ \mathrm{A}[\x_{\textrm{d}}]\mathrm{B}[\x_{\textrm{q}}]^\tran \big]$ is given in Appendix~\ref{sec:App A}. Finally, the covariance matrix of $\tilde{\n}$ is obtained as
\begin{align}
    \mathbf{\Sigma}_{\tilde{\n}} = \C_{\tilde{\n}} - \boldsymbol{\mu}_{\tilde{\n}}\boldsymbol{\mu}_{\tilde{\n}}^\tran. \label{eq:Sigma_n'}
\end{align}

\section{Formulation of Two-Stage hoML} \label{sec:App C}

In the first stage of \ac{two-stage hoML}, we estimate $\x_{\textrm{q}}$ in \eqref{eq:xq} (i.e., the output of the 1-bit \acp{DAC}) by means of the homotopy algorithm in \cite{Sha21}, which approximates a binary \ac{ML} detection problem by solving a sequence of penalty-based non-convex relaxations. In the second stage, we apply full-search \ac{ML}-based data detection to this estimate to detect the data symbol vector $\u$. Define $\tilde{\u} \triangleq \big[ \Re[\u]^\tran, \Im[\u]^\tran\big]^\tran \in \mathbb R^{2K}$ and
\begin{align}
    \tilde{\x}_{\textrm{q}} & \triangleq \big[ \Re[\x_{\textrm{q}}]^\tran, \Im[\x_{\textrm{q}}]^\tran\big]^\tran = [\tilde{x}_{\textrm{q},1}, \ldots, \tilde{x}_{\textrm{q},2N}] \in \mathbb R^{2N}, \\
    \tilde{\W} & \triangleq \begin{bmatrix}
        \Re [\W] & -\Im[\W]\\
        \Im [\W] & \Re[\W]
    \end{bmatrix} ^\tran = [\tilde{\w}_{1}, \ldots, \tilde{\w}_{2N}]
    \in \mathbb R^{2K \times 2N}.
\end{align}

In the first stage, the penalty-based \ac{ML} detection problem of the binary signal $\x_{\textrm{q}}$ based on either $\y$ in \eqref{eq:y} or $\r$ in \eqref{eq:r} is formulated as \cite{Sha21}
\begin{align}\label{eq:bML}
      \underset{\substack{-\sqrt{\frac{\eta_{\rx}}{2}}\leq\tilde{x}_{\textrm{q},n}\leq\sqrt{\frac{\eta_{\rx}}{2}},\\ \forall n \in \{1,\dots,2N\}}}{\minimize}& \ \zeta_{\textrm{bML}}(\tilde{\x}_{\textrm{q}}) - \lambda \|\tilde{\x}_{\textrm{q}}\|_2^2, 
\end{align}
with $\lambda \ge 0$ and where $\zeta_{\textrm{bML}}(\tilde{\x}_{\textrm{q}})$ is the log-likelihood function of $\tilde{\y}$ or $\tilde{\r}$, as given in \cite{Sha21}. For a given $\lambda>0$, the penalty term $-\lambda\|\tilde{\x}_{\textrm{q}}\|_2^2$ pushes the entries of $\tilde{\x}_{\textrm{q}}$ toward the boundaries $\pm\sqrt{\frac{\eta_{\rx}}{2}}$. Then, the homotopy algorithm in \cite{Sha21} gradually increases $\lambda$ and finally returns the binary signal $\tilde{\x}_{\textrm{q}}^\star \triangleq [\tilde{x}_{\textrm{q},1}^\star, \ldots, \tilde{x}_{\textrm{q},2N}^\star] \in \mathbb R^{2N}$ as an estimate of $\tilde{\x}_{\textrm{q}}$. In the second stage, following similar steps as in \eqref{eq:L(x)}--\eqref{eq:Doubly ML}, the \ac{ML} data detection problem applied to $\tilde{\x}_{\textrm{q}}^\star$ is formulated as
\begin{align}\label{eq:hoML}
    \hspace{-2mm} \hat{\u}_{\textrm{two-stage hoML}} = \argmax_{\u \in \setS^K} \sum_{n=1}^{2N} \mathrm{log} \bigg(\Phi\bigg(\sqrt{\frac{2}{\sigma^2}}\tilde{x}_{\textrm{q},n}^\star\tilde{\w}_n^\tran\tilde{\u}\bigg)\bigg).
\end{align}
The \ac{two-stage hoML} method has computational complexity of $\mathcal{O}(I_{\textrm{hom}} N^2 + N L^K)$, where $I_{\textrm{hom}}$ is the number of homotopy iterations in the first stage.

\addcontentsline{toc}{chapter}{References}
\bibliographystyle{IEEEtran}
\bibliography{refs_abbr,refs}

\end{document}